\documentclass[traditabstract,letter,usenatbib,longauth]{aa} % for the letters 
\usepackage{longtable,lscape}
\usepackage{rotating}
\usepackage{subfigure}
\usepackage{caption}
\usepackage{xcolor}
\usepackage{graphicx}

\usepackage{natbib}
    \usepackage{savesym}
    \usepackage{amsmath}
    \savesymbol{iint}
    \usepackage[varg]{txfonts}
    \restoresymbol{TXF}{iint}

\usepackage{amsmath}
\newcommand{\hii}{\relax \ifmmode {\mbox H\,{\scshape ii}}\else H\,{\scshape ii}\fi}
\newcommand{\mi}{\relax \ifmmode {\mu{\mbox m}}\else $\mu$m\fi}
\newcommand{\ha}{\relax \ifmmode {\mbox H}\alpha\else H$\alpha$\fi}
\newcommand{\hb}{\relax \ifmmode {\mbox H}\beta\else H$\beta$\fi}
\newcommand{\hg}{\relax \ifmmode {\mbox H}\beta\else H$\gamma$\fi}
\newcommand{\hd}{\relax \ifmmode {\mbox H}\beta\else H$\delta$\fi}

\newcommand{\sii}{\relax \ifmmode {\mbox S\,{\scshape ii}}\else S\,{\scshape ii}\fi}
\newcommand{\siii}{\relax \ifmmode {\mbox S\,{\scshape iii}}\else S\,{\scshape iii}\fi}
\newcommand{\nii}{\relax \ifmmode {\mbox N\,{\scshape ii}}\else N\,{\scshape ii}\fi}
\newcommand{\oii}{\relax \ifmmode {\mbox O\,{\scshape ii}}\else O\,{\scshape ii}\fi}
\newcommand{\oiii}{\relax \ifmmode {\mbox O\,{\scshape iii}}\else O\,{\scshape iii}\fi}
\newcommand{\neiii}{\relax \ifmmode {\mbox Ne\,{\scshape iii}}\else Ne\,{\scshape iii}\fi}
 
 \newcommand{\rdostres}{\relax \ifmmode {\,\mbox{R}}_{\rm 23}\else \,\mbox{R}$_{\rm 23}$\fi}

\newcommand{\NDOS}{\mbox{\it N2}}

\begin{document}
   \title{Discovering extremely compact and metal-poor, star-forming dwarf galaxies out to  $z\sim$\,0.9 in the VIMOS Ultra-Deep Survey\thanks{Based on data obtained with the European Southern Observatory Very Large Telescope, Paranal, Chile, under Large Program \mbox{185.A-0791}.}%\fnmsep\thanks{Tables 1 and 2 are only available in electronic form at the CDS via http://cdsweb.u-strasbg.fr/cgi-bin/qcat?J/A+A/}
   }
%   \subtitle{II. Evolutionary insights from relations between mass, metallicity and SFR}
\author{R. Amor\'in\inst{1} %\fnmsep\thanks{Based on data obtained with the European Southern Observatory Very Large Telescope, Paranal, Chile, under Large Program 185.A-0791.}
\and V. Sommariva\inst{5,1}
\and M. Castellano\inst{1}
\and A. Grazian\inst{1}
\and L. A. M. Tasca\inst{2}
\and A. Fontana\inst{1}
\and L. Pentericci\inst{1}
%VUDS BUILDERS
\and P. Cassata\inst{2}
\and B. Garilli\inst{4}
\and V. Le Brun\inst{2}
\and O. Le F\`evre\inst{2}
\and D. Maccagni\inst{4}
%\and L. Pentericci\inst{4}
\and R. Thomas\inst{2}
\and E. Vanzella\inst{3}
%\and L. A. M. Tasca\inst{1}
\and G. Zamorani \inst{3}
\and E. Zucca\inst{3}
%VUDS MEMBERS:
%\and R. Amorin\inst{4}
\and S. Bardelli\inst{3}
\and P. Capak\inst{12}
\and L. P. Cassar\'a\inst{4}
%\and M. Castellano\inst{4}
\and A. Cimatti\inst{5}
\and J.G. Cuby\inst{2}
\and O. Cucciati\inst{5,3}
\and S. de la Torre\inst{2}
\and A. Durkalec\inst{2}
%\and A. Fontana\inst{4}
\and M. Giavalisco\inst{13}
%\and A. Grazian\inst{4}
\and N. P. Hathi\inst{2}
\and O. Ilbert\inst{2}
\and B. C. Lemaux \inst{2}
\and C. Moreau\inst{2}
\and S. Paltani\inst{9}
\and B. Ribeiro\inst{2}
\and M. Salvato\inst{14}
\and D. Schaerer\inst{10,8}
\and M. Scodeggio\inst{4}
%\and V. Sommariva\inst{5,1}
\and M. Talia\inst{5}
\and Y. Taniguchi\inst{15}
\and L. Tresse\inst{2}
\and D. Vergani\inst{6,3}
\and P.W. Wang\inst{2}
%- VUDS associates
\and S. Charlot\inst{7}
\and T. Contini\inst{8}
\and S. Fotopoulou\inst{9}
\and C. L\'opez-Sanjuan\inst{11}
\and Y. Mellier\inst{7}
\and N. Scoville\inst{12}
}
%\offprints{R. Amor\'in \email{ricardo.amorin@oa-roma.inaf.it}}
   \institute{INAF--Osservatorio Astronomico di Roma, via di Frascati 33, I-00040,  Monte Porzio Catone, Italy
\and
%1
Aix Marseille Universit\'e, CNRS, LAM (Laboratoire d'Astrophysique  de Marseille) UMR 7326, 13388, Marseille, France
\and
%2
INAF--Osservatorio Astronomico di Bologna, via Ranzani,1, I-40127, Bologna, Italy
\and
%3
INAF--IASF, via Bassini 15, I-20133,  Milano, Italy
\and
%5
University of Bologna, Department of Physics and Astronomy (DIFA), V.le Berti Pichat, 6/2 - 40127, Bologna
\and
%6
INAF--IASF Bologna, via Gobetti 101, I--40129,  Bologna, Italy
\and
%7
Institut d'Astrophysique de Paris, UMR7095 CNRS, 
Universit\'e Pierre et Marie Curie, 98 bis Boulevard Arago, 75014, Paris, France
\and
%8
Institut de Recherche en Astrophysique et Plan\'etologie - IRAP, CNRS,
Université de Toulouse, UPS-OMP, 14, avenue E. Belin, F31400, Toulouse, France
\and
%9
Department of Astronomy, University of Geneva, ch. d'Écogia 16, CH-1290 Versoix
\and
%10
Geneva Observatory, University of Geneva, ch. des Maillettes 51, CH-1290 Versoix, Switzerland
\and
%11
Centro de Estudios de F\'isica del Cosmos de Arag\'on, Teruel, Spain
\and
%12
Department of Astronomy, California Institute of Technology, 1200 E. California Blvd., MC 249-17, Pasadena, CA 91125, USA
\and
%13
Astronomy Department, University of Massachusetts, Amherst, MA 01003, USA
\and
%14
Max-Planck-Institut f\"ur Extraterrestrische Physik, Postfach 1312, D-85741, Garching bei M\"unchen, Germany
\and
%15
Research Center for Space and Cosmic Evolution, Ehime University, Bunkyo-cho 2-5, Matsuyama 790-8577, Japan
              }
             
   \date{}  

 \abstract
{We report the discovery of 31 low-luminosity ($-14.5$\,$\ga$\,$M_{\rm AB}(B)$\,$\ga$\,$-18.8$),  
extreme emission line galaxies (EELGs) at $0.2\la$\,$z$\,$\la 0.9$
identified by their unusually high rest-frame equivalent widths 
(100$\leq$\,EW[\oiii]\,$\leq$1700\AA) as part of 
%an initial phase ($\sim$\,40\%) of 
the VIMOS Ultra Deep Survey (VUDS). 
VIMOS optical spectra of unprecedented sensitivity ($I_{\rm AB}$\,$\sim$\,25\,mag) 
along with multiwavelength photometry and HST imaging are used to 
investigate spectrophotometric properties of this unique 
sample and to explore, for the first time, the very low stellar mass end 
(M$_{\star}$\,$\la$\,10$^{8}$\,M$_{\odot}$) 
of the luminosity-metallicity (LZR) and mass-metallicity (MZR) 
relations at $z<1$. 
Characterized by their extreme compactness ($R_{50} <$\,1 kpc),  
low stellar mass and enhanced specific star formation rates  
(sSFR\,$=$\,SFR/M$_{\star}$\,$\sim$\,10$^{-9}$-10$^{-7}$yr$^{-1}$), the VUDS 
EELGs are blue dwarf galaxies likely experiencing the first stages 
of a vigorous galaxy-wide starburst. 
Using $T_{e}$-sensitive \textit{direct} and strong-line methods, we find 
that VUDS EELGs are low-metallicity (7.5$\la$12$+\log$(O/H)$\la$\,8.3) 
galaxies with high ionization conditions ($\log(q_{\rm ion})$\,$\ga$\,8\,cm\,s$^{-1}$), 
%(\oiii/\oii\,$=4\pm5$), 
including at least three EELGs showing He{\sc ii}$\lambda$\,4686\AA \ 
emission and four extremely metal-poor ($\la$10\% solar) galaxies. 
The LZR and MZR followed by VUDS EELGs show relatively large scatter,
being broadly consistent with the extrapolation toward low luminosity 
and mass from previous studies at similar redshift. 
However, we find evidence that galaxies with younger and
more vigorous star formation -- as characterized by their larger % \hb \ and [\oiii] 
EWs, ionization and sSFR -- tend to be more metal poor at a given stellar mass.
}
% Results (mandatory)
%{.}
% Conclusions (optional)
%{Results from this pilot study underlines the powerful of VUDS spectroscopy...} 
% 5 {} token are mandatory


\titlerunning {Low-metallicity starbursts in VUDS}        
\authorrunning{R. Amor\'in et al.}
  \maketitle
 
%
%________________________________________________________________

\section{Introduction}
\label{sect:intro}

%The stellar mass build-up and subsequent chemical evolution 
%of low-mass galaxies are mostly driven by episodic 
%galaxy-wide starburst episodes \citep[e.g.][]{Stinson2007}.  

Over the last 8 billion years % , when more massive galaxies have
                              % already formed their stellar
                              % populations and the cosmic star
                              % formation history show a strong
                              % decline \citep{Hopkins2006}, 
a large fraction of low-mass  (M$_{\star}\la$\,10$^{9}$\,M$_{\odot}$) 
galaxies are still seen rapidly assembling most of their 
present-day stellar mass \citep{Cowie1996,Perez-Gonzalez2008}.
Tracing the spectrophotometric properties of these vigorous 
star-forming \textit{dwarf} galaxies (SFDGs) out to $z\sim$1 is 
essential not only to study how they evolve through cosmic 
time, but also to understand the physical mechanisms driving 
the first stages of stellar mass build-up and chemical enrichment.  
To this end, key insights %of these mechanisms 
can be obtained from the tight relations found between 
stellar mass, metallicity and star formation rate (SFR). 
%, which constitute a record of the integral star formation history of galaxies.  
However, the shape and normalization of these relations at 
different redshifts are still poorly constrained at their low-mass end. 
While in the local Universe the mass-metallicity relation 
(MZR) has been extended down to 10$^8$\,M$_{\odot}$
\citep[e.g.][]{Andrews2013}, at intermediate and high 
redshifts, dwarf galaxies are strongly underrepresented 
\citep[e.g.][]{Henry2013}. 

These SFDGs are usually identified by their blue colors, high
surface brightness and strong emission-lines. 
They include a rare population of extreme emission-line 
galaxies (EELGs) with the largest nebular content and 
lowest metal abundances 
\citep[e.g.][]{Kniazev2004,Papaderos2008,Hu2009,Atek2011,Morales-Luis2011}.
Due to their high equivalent widths (EWs), an increasing number 
of EELGs are being discovered and characterized by deep 
spectroscopic surveys out to $z\sim$1 
\citep[e.g.][]{Hoyos2005,Ly2014,Amorin2014a} and 
beyond \citep[e.g.][]{vdWel2011,Maseda2014}. 
In this \textit{Letter} we report the discovery of a sample 
of 31 EELGs at $0.2\la z \la 0.9$ identified from the 
\textit{VIMOS Ultra-Deep Survey} \citep[VUDS;][]{LeFevre2014}. 
We study their physical properties as part of a larger, ongoing 
study aimed at investigating the evolution of SFDGs out to 
$z\sim$\,1 using very deep spectroscopy \citep[e.g.][]{Amorin2014a}. 
The sensitivity of our VUDS spectra, detecting emission lines as faint as 
$\sim$1.5$\times$\,10$^{-18}$\,erg s$^{-1}$ cm$^{-2}$, %(3$\sigma$), 
makes it possible e.g., to derive $T_e$-based metallicities 
for a fraction of such faint galaxies. 
Thus, the present sample extends previous studies 
of star-forming (SF) galaxies at similar redshifts in size and limiting 
magnitude \citep{Henry2013,Ly2014}, allowing us to study in 
greater detail the LZR and MZR at $z<1$ two decades below 
10$^{9}$M$_{\odot}$ with galaxies showing a 
wide range of properties, including a number 
of extremely metal-poor galaxies. % ($Z<0.1Z_{\odot}$). 
Throughout this paper we adopt a standard $\Lambda$-CDM 
cosmology with $h$ = 0.7, $\Omega_m$ = 0.3 and 
$\Omega_\Lambda$ = 0.7.

\section{Observations and sample selection} 
\label{sect:observations}
%-------------------------------------------------------------
  \begin{figure}[t]
\centering
   \includegraphics[angle=0,width=9.cm]{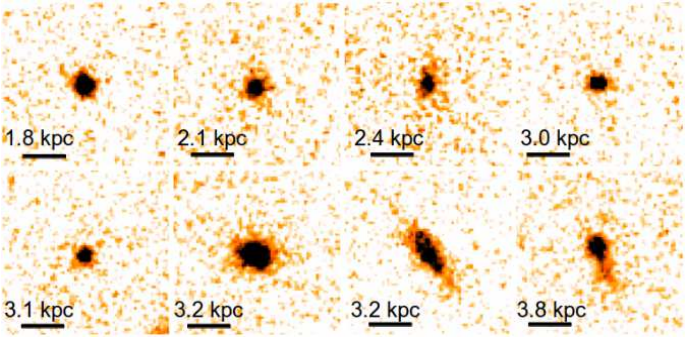} %\hspace{5pt}
  \caption{HST $F814W$-band imagery of EELGs in the COSMOS and ECDF fields covered by VUDS. Each postage stamp is 2$^{"}$ on a side. 
  }\label{morphology}
   \end{figure}
%-------------------------------------------------------------
The \textit{Vimos Ultra Deep Survey} is a deep spectroscopic legacy 
survey of $\sim$\,10$^4$ galaxies carried out using VIMOS at 
ESO-VLT \citep{LeFevre2003}. 
This survey is aimed at providing a complete census of the 
SF galaxy population at $2 \la z \la 7$, covering $\sim$\,1\,deg$^2$ 
in three fields: COSMOS, ECDFS, and VVDS-2h. 
The VIMOS spectra consist of 14h integrations in the LRBLUE 
and LRRED grism settings, covering a combined 
wavelength range $3650<\lambda<9350$\AA, with a spectral resolution 
R$\sim$230. Data reduction, redshift measurement, and assessment 
of the reliability flags are described in detail in the survey 
and data presentation paper \citep{LeFevre2014}.  

The targets of VUDS have been primarily selected to have photometric 
redshifts $z_p > 2.4$ for either of the primary and secondary peaks 
of the PDF. A number of random targets purely magnitude selected to 
$23 \leq I_{\rm AB} \leq 25$ have been added to 
fill empty areas on observed slit masks. As a consequence, we identify a number of
targets with spectroscopic redshift $z_s <2$. 
%VUDS targets have been selected to have photometric redshifts $z_p > 2$. 
%However, additional faint targets (23.5$<$\,$I_{\rm AB}$\,$<$25.5) 
%have been randomly selected to fill the VIMOS masks \citep[see][]{LeFevre2014}. 
%We thus identify a number of them with spectroscopic redshift $z_s <2$.
Many of these targets are galaxies with prominent optical emission 
lines, such as [\oii]$\lambda$\,3727 or [\oiii]$\lambda$\,5007, 
that artificially boost the observed magnitudes in the stellar 
spectral energy distributions (SED). 

For this \textit{Letter} a representative sample of 31 EELGs 
(12 from COSMOS, 11 from VVDS-2h, and 8 from ECDFS) 
with mean $I_{\rm AB}$\,$\sim$24.5 mag was identified from an early 
version of VUDS data containing $\sim$40\% of the final sample. 
We first consider primary and secondary target galaxies with 
very reliable spectroscopic redshift (98\% and 100\% confidence level), 
%i.e., flagged as 3, 4, 23 and 24, respectively, 
at $z\leq 0.93$.  
%\citep[e.g.][]{Atek2011,Amorin2014a},  
We then select galaxies with at least three 
emission lines detected, [\oii], [\oiii], and \hb, 
and EW[\oiii]$>$\,100\AA. 
The first criterion ensures the derivation of gas-phase 
metallicities and the second allows us to select 
SFDGs with the highest \textit{specific} SFR 
\citep[e.g.][]{Atek2011,Ly2014,Amorin2014a}. 

While our EELGs look unresolved in ground-based images 
precluding a full morphological analysis, 
morphological information can be obtained for a subset 
of 16 EELGs that have been observed by the HST-ACS in the 
$F814W$ ($I$) band. As illustrated in Fig.~\ref{morphology}, 
EELGs include galaxies with both round and irregular shapes,  
showing angular sizes $<$\,1$^{"}$. Using the automated method 
presented in \citet{Tasca2009} for the EELGs imaged by 
the ACS we derive circularized half-light radii, 
$r_{50}$\,$=$\,$R_{50}$\,$(b/a)^{0.5}$\,$\sim$\,0.4-0.8 kpc 
% \footnote{We have circularized the half-light radii as $r_{50}$$=$$R_{50}$\,$q^{0.5}$, where $q$ is the axial ratio, $b/a$.} 
%0.4\,$\la$\,$r_{50}$\,$\la$\,0.8 kpc, 
thus confirming their extreme compactness. 
In most cases, we find these EELGs with no clear signs of ongoing 
mergers or very close companions.
%-------------------------------------------------------------
%  \begin{figure}[t!]
%   \includegraphics[angle=0,width=8.5cm]{fig2.eps} %\hspace*{0.5cm}
%   %\includegraphics[angle=0,width=8.7cm]{./figures_paper/spec2.eps} 
%\hspace{5pt}
%  \label{spectrum}\caption{Deep VIMOS spectrum of a high ionization,
%    (EW([\oiii])$\sim$\,1800\AA)
% extremely metal-poor ($Z$\,$\sim$\,0.07\,$Z_{\odot}$) EELG in VUDS. 
%  A zoomed version is shown in the bottom panel. 
%  The main nebular emission lines are labeled. 
%  }
%   \end{figure}
%-------------------------------------------------------------  

\section{Physical properties of VUDS EELGs}
\label{sect:analysis}

Deep VUDS spectra for the sample of EELGs are presented 
in Fig.~A.1\footnote{\label{note2}Only available in the electronic edition of the journal}. 
Long exposure times allow us to detect in most cases a remarkably faint continuum 
($\sim$\,5$\times$\,10$^{-20}$erg\,s$^{-1}$\,cm$^{2}$\,\AA$^{-1}$, 1$\sigma$) 
and very faint lines, such as [\oiii]$\lambda$4363 or [\nii]$\lambda$6584. 
In Table~1\footnote{\label{note1}Tables 1 and 2 are only available in electronic form at the CDS via http://cdsweb.u-strasbg.fr/cgi-bin/qcat?J/A+A/} %\footnotemark[\ref{note1}] 
we present line fluxes and uncertainties for the most relevant detected emission lines, which were 
performed manually using the IRAF task {\sl splot} following  \citet{Amorin2012}. 
Reddening corrections were performed using the Balmer 
decrement, whenever available, and adopting the \citet{Calzetti2000} 
extinction law. 
For those EELGs with \ha/\hb \ or \hb/\hg\ measurements the median reddening 
is $E(B-V)^{\rm med}_{\rm gas}$\,$=$\,0.26 ($\sigma=0.14$), in good agreement
with previous studies for EELGs \citep[e.g.,][]{Dominguez2013,Ly2014,Amorin2014a}. 
In those cases where $E(B-V)_{\rm gas}$ cannot be measured through 
\ha/\hb \ or \hb/\hg, or where its values are smaller than the theoretical values 
for the Case B recombination ($T_e=$\,2$\times$10$^4$K, $n_e=$100\,cm$^{-3}$), we adopt $E(B-V)_{\rm gas}$\,$=$\,$E(B-V)_{\star}$, where $E(B-V)_{\star}$ is the stellar extinction derived from the SED fitting described in Section\,3.2. 
This assumption seems reasonable since median values of stellar 
($E(B-V)^{\rm med}_{\star}$\,$=$0.25, $\sigma=0.14$) and 
gas extinctions are in excellent agreement for galaxies for which 
both values are available. The adopted reddening constant for each galaxy 
is listed in Table~1. 
In the following sections we describe the derivation of the main 
physical properties for the EELG sample, which are presented 
in Table~2$^2$. % \footnotemark[\ref{note1}].

\subsection{Ionization and metallicity properties from VUDS spectra}
%-------------------------------------------------------------
   \begin{figure}[t!]
   \centering
%\begin{minipage}{90mm}
  \includegraphics[angle=90,width=9.cm]{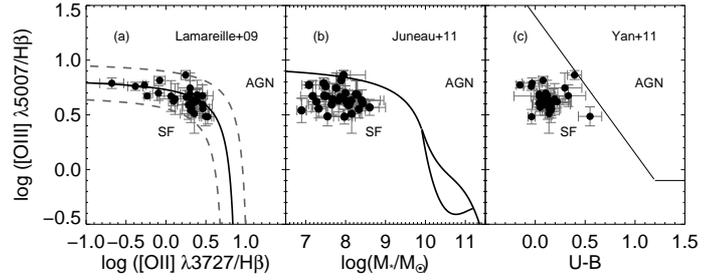} 
   \caption{Diagnostic diagrams. Lines show the empirical separations 
   between SF galaxies and AGNs. 
    } 
%\end{minipage}
   \label{diagnostics}
   \end{figure}
%-------------------------------------------------------------

In Figure~\ref{diagnostics} we study the ionization properties of the 
EELG sample using three diagnostic diagrams based on strong 
emission line ratios. 
Our sample galaxies populate the region of SF galaxies with 
the highest excitation ([\oiii]/\hb$\sim$\,5$\pm$2).  
Consistently with their low masses and blue $U-B$ colors, none 
of them shows indication of an Active Galactic Nuclei (AGN) activity. %, as shown in Fig.~\ref{diagnostics}. 
%, e.g., X-ray emission (TO BE CHECKED), very broad Balmer lines and/or
%hard ionization lines such as [Ne\,{\sc v}]. 
Our EELGs, however, are located near the limits between SF and AGN 
regions in Fig.~\ref{diagnostics} due to their high  
ionization conditions, as suggested by their high [\oiii]/[\oii] 
ratios (Fig.~\ref{diagnostics}$a$). 
In the most extreme case, [\oiii] shows EW of
$\sim$\,1700\AA, while the [\oii] line is only barely detected. 
Moreover, in three EELGs, we tentatively 
detect ($\sim$2.5$\sigma$) He{\sc ii}$\lambda$\,4686\AA\ 
emission, suggesting the presence of very young, hot stars. 
Being rare at $z<1$ 
\citep[e.g.,][]{Jaskot2013,Nakajima2013,Amorin2014a},
these EELGs show ionization parameters 
($\log(q_{\rm ion})$\,$\ga$\,8\,cm\,s$^{-1}$) 
comparable to some low-luminosity high redshift galaxies 
\citep[e.g.,][]{Fosbury2003,Amorin2014b}. 

In seven EELGs we detect ($\geq$\,2$\sigma$) the intrinsically faint 
$T_{e}-$sensitive auroral line [\oiii]$\lambda$4363\AA. 
For these galaxies we derive metallicity using the direct 
method \citep{Hagele2008}. 
In addition, we derive metallicities for the entire sample using the 
$R23(\equiv$(\oii$+$\oiii)/\hb) parameter and the calibration 
of %\citet{Kobulnicky2003} based on the photoionisation models by 
\citet{McGaugh1991}. %, which includes additional terms as a function of [\oii]/[\oiii] to account for the known dependence of $R23$ on $q_{\rm ion}$ and effective temperature. Moreover, 
Following \citet[][]{Perez-Montero2013} we apply the linear 
relations detailed in \citet{Lamareille2006} to make these 
R23 metallicities consistent with those derived using the 
direct method.  
In order to break the degeneracy of $R23$ (i.e., to choose between 
the lower or upper branch) we use two additional indicators.  
For EELGs at $z\la$\,0.45 we choose the branch that best matches 
the metallicity obtained from the \NDOS($\equiv$\nii/\ha) parameter 
and the calibration by \citet{Perez-Montero2009}, 
while for EELGs at $z\ga$\,0.45 we choose the branch that best 
matches the metallicity from the calibrations based on the 
[\neiii], [\oii], and [\oiii] line ratios of \citet{Maiolino2008}.  
The difference between direct and strong-line metallicity estimations 
for the seven galaxies with [\oiii]$\lambda$4363\AA\  
is $<$\,0.2 dex. 
We find the metallicity of our EELG spanning a wide range of 
subsolar values (7.5$\la$12$+\log$(O/H)$\la$\,8.3), including 
four extremely metal-poor galaxies ($Z\la$\,0.1$Z_{\odot}$).   

\subsection{Stellar properties from multiwavelength SED fitting}

Stellar masses and rest-frame absolute magnitudes of EELGs 
 were derived by fitting their stellar SEDs.
In short, we %use the \textsl{zphot} code to 
fit \citet{Bruzual2003} stellar population synthesis models to 
the broad-band photometry -- from UV to NIR -- of each galaxy 
using chi-square minimization {following \citet{Castellano2014}. 
Magnitudes are previously corrected from the contribution of prominent optical emission lines following \citet{Amorin2014a}, while models 
assume stellar metallicities that best agree with the observed gas-phase metallicity. }
We adopt a \citet{Chabrier2003} IMF, \citet{Calzetti2000} extinction
law and assume a standard declining exponential star formation 
history. As a result, we find the sample of EELGs in VUDS spanning a  
range of low luminosities, 
$-14.5$\,$\la$\,$M_{\rm AB}(B)$\,$\la$\,$-18.8$, and low 
stellar masses, 6.9$\la$\,$\log($M$_{\star}$/M$_{\odot}) \la$\,8.6. 
%-------------------------------------------------------------
   \begin{figure}[t!]
   \centering
  \includegraphics[angle=0,width=8.3cm]{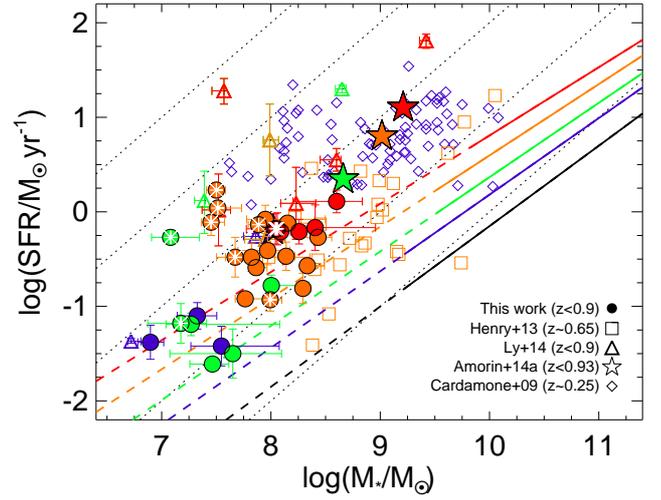}
    \caption{The SFR stellar mass plane of low-mass SFDGs. 
    Solid and dashed lines show the so-called 
    main sequence of galaxies at different redshifts and its extrapolation 
    to low-mass regime, respectively, according to \citet{Whitaker2012}. 
    Dotted lines indicate constant sSFR from 10$^{-10}$\,yr$^{-1}$ (bottom) 
    to 10$^{-6}$\,yr$^{-1}$ (upper).
   Colors indicate redshift bins with mean values
   $\langle$\,$z$\,$\rangle$\,$=$\,0 (black), 
    $\langle$\,$z$\,$\rangle$\,$=$\,0.25 (blue), 
   $\langle$\,$z$\,$\rangle$\,$=$\,0.4 (green), 
    $\langle$\,$z$\,$\rangle$\,$=$\,0.6 (orange) and 
   $\langle$\,$z$\,$\rangle$\,$=$0.8 (red). 
    Asterisks show VUDS EELGs with EW$_{\rm rest}$(\oiii)\,$>$\,200\AA\ and 
    {EW$_{\rm rest}$(\hb)\,$>$\,60\AA.}}
%\end{minipage}
  \label{M-SFR}
   \end{figure}
%-------------------------------------------------------------

\section{The relation between mass, metallicity, and ongoing SFR of 
low-mass galaxies out to $z\sim$1}
\label{sect:discussion}

In Fig.~\ref{M-SFR} we show the SFR-mass diagram for the 
EELGs in VUDS and from the literature. 
Star formation rates are derived from the extinction-corrected 
\ha\ or \hb\ luminosities using the calibration of
\citet{Kennicutt1998} and assuming a \citet{Chabrier2003} IMF. 
At a given redshift, our EELGs show SFRs and stellar masses a 
factor of $\sim$10 lower than similar samples from the literature.  
However, nearly all EELGs shown in Fig.~\ref{M-SFR} are well 
above the extrapolation to low stellar mass of 
the main sequence of galaxies \citep{Whitaker2012} at a given $z$. 
The EELGs in VUDS show enhanced \textit{specific} SFRs 
(sSFR$\sim$10$^{-9}$-10$^{-7}$\,yr$^{-1}$) and SFR surface densities   
(median $\Sigma_{\rm SFR}=$ SFR$/2\pi r^2_{50} =$\,0.35 ($\sigma =$\,0.19)\,M$_{\odot}$\,yr$^{-1}$\,kpc$^{-2}$), 
comparable to more luminous galaxy-wide starbursts at similar and
higher redshifts \citep{Ly2014,Amorin2014a,Amorin2014b}. 
%-------------------------------------------------------------
   \begin{figure}[t!]
   \centering
%\begin{minipage}{90mm}
\includegraphics[angle=0,width=7.6cm]{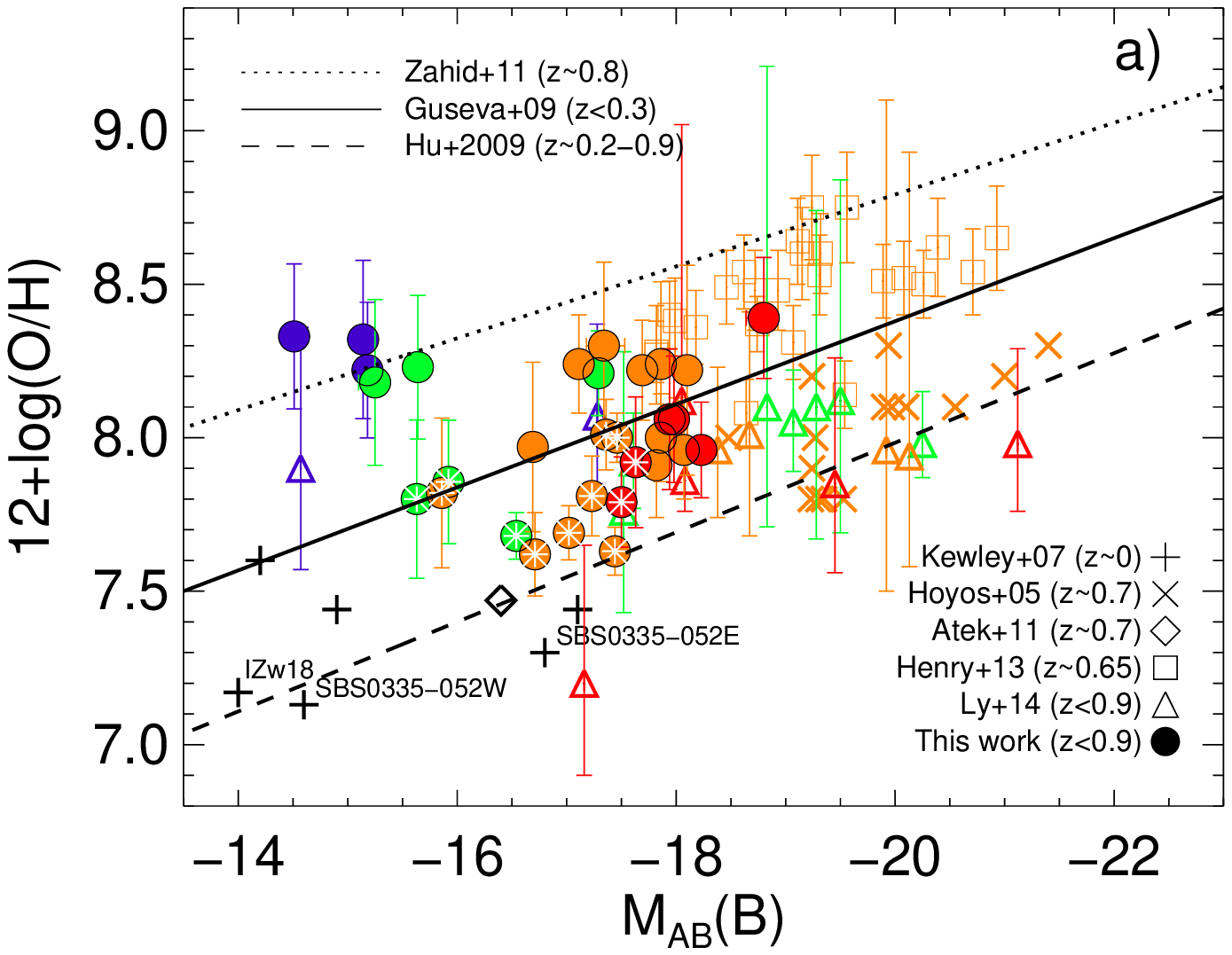}\\\vspace{2mm}
\includegraphics[angle=0,width=7.6cm]{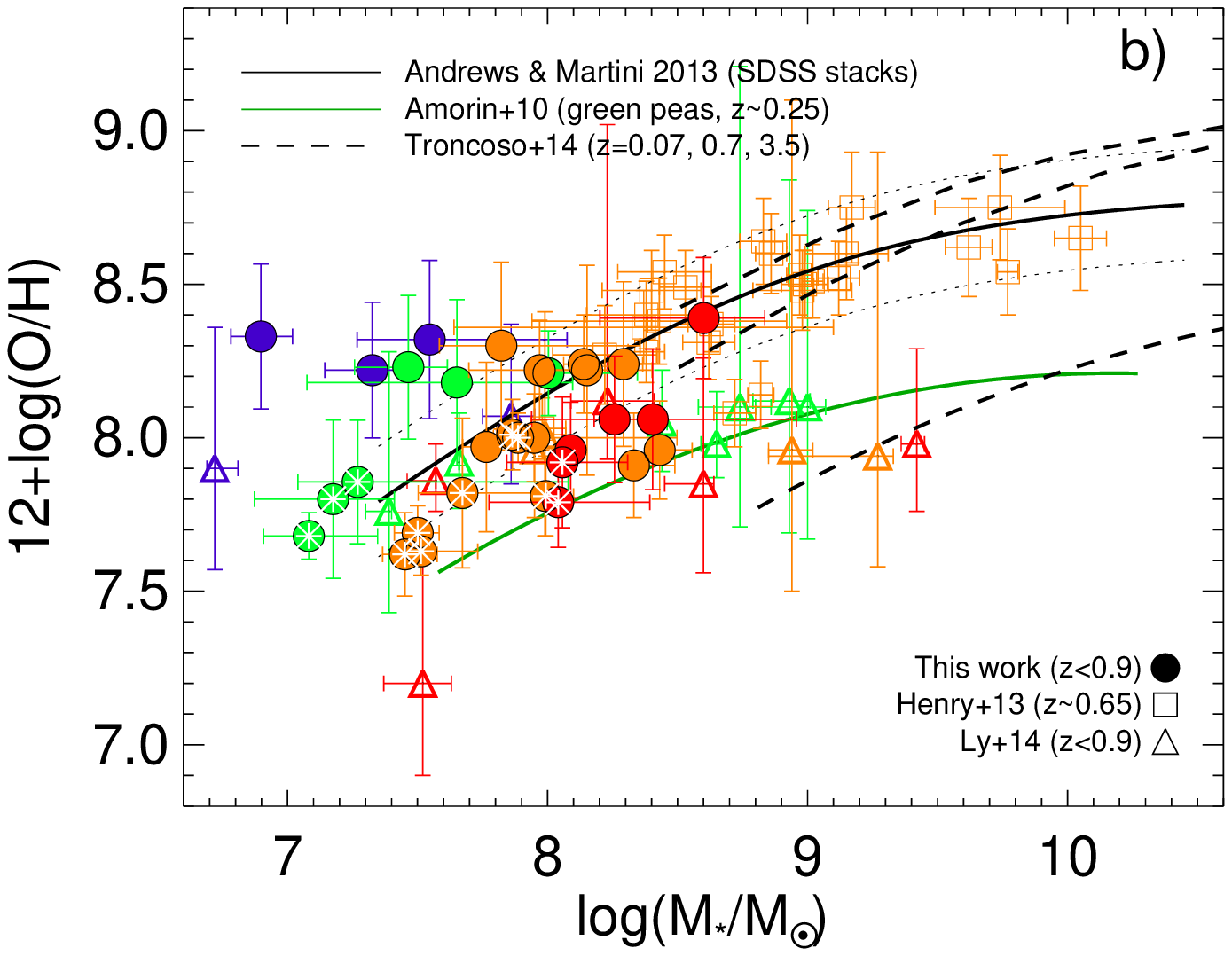}\\\vspace{2mm}
\includegraphics[angle=0,width=7.6cm]{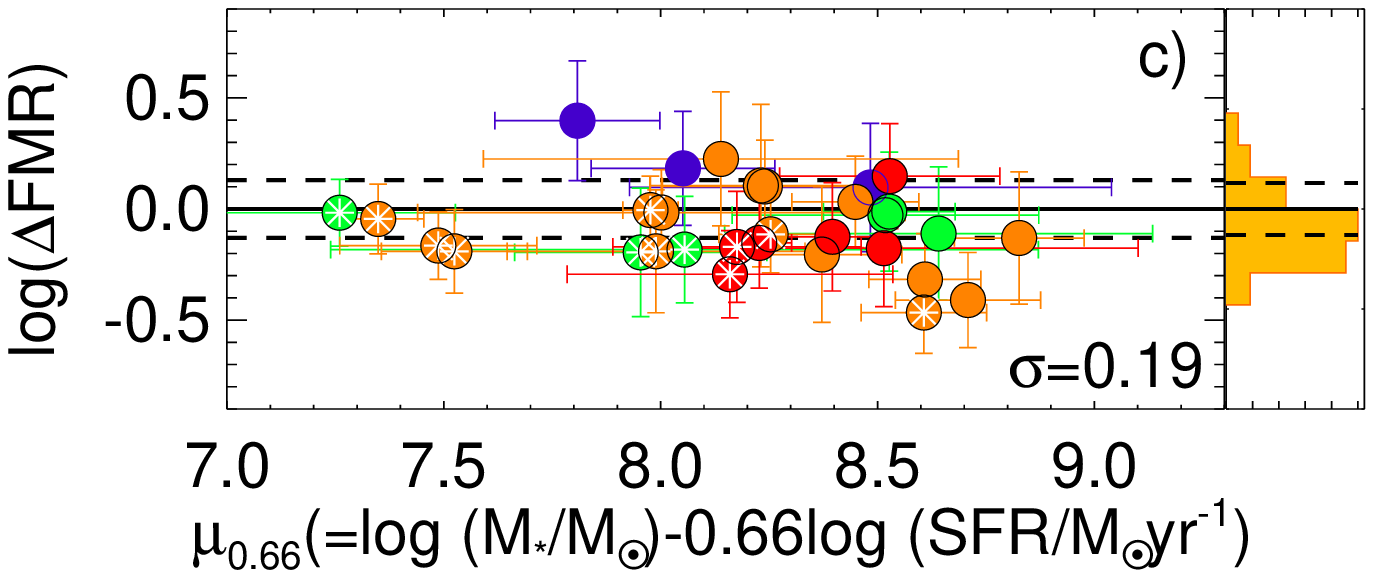}\\\vspace{2mm}
\includegraphics[angle=0,width=7.6cm]{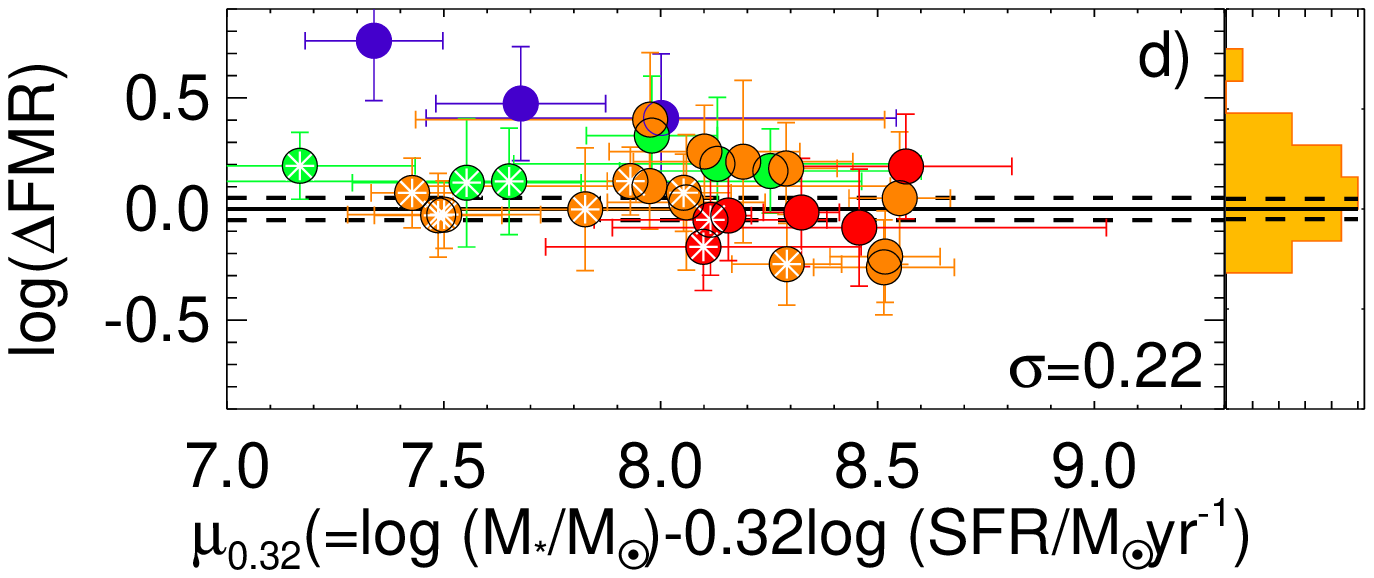}\\
    \caption{{($a$) Luminosity-metallicity and ($b$) mass-metallicity relations for VUDS EELGs and SFDGs from the literature. 
% and the \citet{Maiolino2008} calibration for     metallicities, except for those derived through the direct method. 
     Metallicity differences with respect to the extrapolation to low stellar mass of the FMR by \citet[][]{Andrews2013} and \citet[][]{Mannucci2011} are shown in $(c)$ and $(d)$, respectively. Dashed lines indicate 1$\sigma$ deviations for these relations. %Similar results are obtained using the FMR from \citet{Mannucci2010}.
     Colors and symbols %have the same meaning 
     are as in Fig.~\ref{M-SFR}. 
     The data have been homogenized to the Chabrier IMF and the same 
     strong-line metallicity calibration presented in Section~3.1.} }
%\end{minipage}
   \label{MZR}
   \end{figure}
%-------------------------------------------------------------

In Fig.~\ref{MZR} we study the LZR and MZR traced by 
EELGs in VUDS and other low-mass galaxies at $0<z<1$. 
The EELGs {extend} the LZR down to $M_{\rm AB}(B) \sim -14.5$ and 
the MZR down to M$_{\star} \sim$\,10$^{7}$M$_{\odot}$, which 
means $\sim$1 dex lower than previous studies \citep[e.g.,][]{Henry2013}, thus increasing substantially the number of low-mass galaxies under study, especially at $z$\,$\ga$\,0.5.    
Despite the relatively large scatter, VUDS EELGs appear to 
follow the LZR and MZR of more luminous and massive SFDGs. 
In particular, we find most EELGs in broad agreement with the 
local ($z<0.3$) LZR of \citet{Guseva2009} and MZR of \citet{Andrews2013}, 
which have been derived from galaxies with {$T_e$-based  metallicities. 
There is nevertheless a tendency for EELGs with larger EWs %EW(\hb) and EW(\oiii) 
%(highlighted with asterisks in Fig.~\ref{MZR}) 
to be more metal-poor at a given luminosity, stellar mass, and redshift.} 
These galaxies are those with the highest sSFR, i.e., those with the largest deviations from the main sequence of star formation at a given $z$, shown in Fig.~\ref{M-SFR}. 
While they follow more reliably the LZR traced by extremely 
metal-poor galaxies \citep[e.g.,][]{Kewley2007,Hu2009}, they tend to lie 
below the local MZR, similarly to other extreme galaxies 
\citep[see, e.g., the \textit{green peas}][]{Amorin2010}. 

Part of the above apparent dependence of the MZR on SFR can be 
explained in terms of the fundamental metallicity relation 
\citep[FMR;][]{Mannucci2010}, which suggests that galaxies with higher
sSFR tend to be more metal-poor at a given stellar mass. 
 As shown in Fig.~\ref{MZR}, the position of the VUDS EELGs appears 
broadly consistent with the extrapolation to low masses of the FMR,
independently of the parametrization and metallicity scale adopted. 
We notice, however, that the scatter in the FMR for EELGs 
($\sigma \sim$\,0.20) seems slightly larger than expected for
magnitude-selected samples in the local universe. 

Overall, the above results are consistent with a picture where the
most extreme SFDGs are very gas-rich galaxies experiencing 
an early stage of a galaxy-wide starburst, possibly fed by recent 
accretion of metal-poor gas \citep[e.g.,][]{Amorin2010,SanchezAlmeida2014}. 
In this picture, at least part of the scatter in the above scaling relations 
could be produced by differences in the accretion and star formation histories. 

Figure~\ref{MZR} also suggests that the shape of the MZR can be very 
sensitive to selection effects in its very low-mass end. 
Gas-rich dwarfs with prominent emission lines, enhanced sSFR and low 
metallicities may be overrepresented with respect to the global 
population of SFDGs in magnitude-selected spectroscopic samples at these 
redshifts, making the shape of the MZR at low mass not entirely 
representative of main sequence galaxies.  
Clearly, a thorough study using the deepest spectroscopy available 
for a statistical significant complete sample of SFDGs is much needed 
to test this hypothesis. 
Forthcoming analysis of VUDS galaxies at $z<1$ using 
the complete database will enable us to scrutinize in detail the 
underexplored low-mass universe at $z<1$.  

\begin{acknowledgements}
We thank the anonymous referee for helpful comments that helped to 
improve this manuscript.
We thank the ESO staff for their continuous support for the VUDS 
survey, particularly the Paranal staff conducting the observations and 
Marina Rejkuba and the ESO user support group in Garching.
This work is supported by funding from the European Research Council Advanced 
Grant ERC-2010-AdG-268107-EARLY and by INAF Grants PRIN 2010, PRIN 2012 
and PICS 2013. 
RA and AF acknowledge the FP7 SPACE project “ASTRODEEP” (Ref.No: 312725), 
supported by the European Commission. 
AC, OC, MT and VS acknowledge the grant MIUR PRIN 2010--2011.  
DM gratefully acknowledges LAM hospitality during the initial phases of the project. 
This work is based on data products made available at the CESAM data center, 
Laboratoire d'Astrophysique de Marseille. 
This work partly uses observations obtained with MegaPrime/MegaCam, a joint project 
of CFHT and CEA/DAPNIA, at the Canada-France-Hawaii Telescope (CFHT) which 
is operated by the National Research Council (NRC) of Canada, the Institut National 
des Sciences de l'Univers of the Centre National de la Recherche Scientifique (CNRS) 
of France, and the University of Hawaii. This work is based in part on data products 
produced at TERAPIX and the Canadian Astronomy Data Centre as part of the 
Canada-France-Hawaii Telescope Legacy Survey, a collaborative project of NRC 
and CNRS.

\end{acknowledgements}

%\begin{thebibliography}{aa}
\bibliographystyle{aa}
\bibliography{•}

\begin{thebibliography}{132}
\expandafter\ifx\csname natexlab\endcsname\relax\def\natexlab#1{#1}\fi

%\bibitem[Allende Prieto et al.(2001)]{AllendePrieto2001} Allende Prieto, C., Lambert, D.~L., \& Asplund, M.\ 2001, \apjl, 556, L63 

\bibitem[{{Amor{\'{\i}}n} {et~al.}(2010){Amor{\'{\i}}n}, {P{\'e}rez-Montero},
  \& {V{\'{\i}}lchez}}]{Amorin2010} {Amor{\'{\i}}n}, R.~O., {P{\'e}rez-Montero}, E., \& {V{\'{\i}}lchez}, J.~M.  2010, \apjl, 715, L128

\bibitem[Amor{\'{\i}}n et al.(2012)]{Amorin2012} Amor{\'{\i}}n, R., P{\'e}rez-Montero, E., V{\'{\i}}lchez, J.~M., \& Papaderos, P.\ 2012, \apj, 749, 185 

%\bibitem[Amor{\'{\i}}n et al.(2012b)]{Amorin2012b} Amor{\'{\i}}n, R., V{\'{\i}}lchez, J.~M., H{\"a}gele, G.~F., et al.\ 2012, \apjl, 754, L22 

\bibitem[Amor{\'{\i}}n et al.(2014a)]{Amorin2014a} Amor{\'{\i}}n, 
R., P{\'e}rez-Montero, E., Contini, T., et al.\ 2014a, arXiv:1403.3441 

\bibitem[Amor{\'{\i}}n et al.(2014b)]{Amorin2014b} Amor{\'{\i}}n, 
R., Grazian, A., Castellano, M., et al.\ 2014b, \apjl, 788, L4 

\bibitem[Andrews \& Martini(2013)]{Andrews2013} Andrews, B.~H., \& Martini, P.\ 2013, \apj, 765, 140 

\bibitem[Atek et al.(2011)]{Atek2011} Atek, H., Siana, B., Scarlata, C., et al.\ 2011, \apj, 743, 121 

%\bibitem[Baldwin et al.(1981)]{Baldwin1981} Baldwin, J.~A., Phillips, M.~M., \& Terlevich, R.\ 1981, \pasp, 93, 5 

%\bibitem[Brammer et al.(2012)]{Brammer2012} Brammer, G.~B., S{\'a}nchez-Janssen, R., Labb{\'e}, I., et al.\ 2012, \apjl, 758, L17 

%\bibitem[Brammer et al.(2013)]{Brammer2013} Brammer, G.~B., van Dokkum, P.~G., Illingworth, G.~D., et al.\ 2013, \apjl, 765, L2 

\bibitem[{{Bruzual} \& {Charlot}(2003)}]{Bruzual2003}{Bruzual}, G. \& {Charlot}, S. 2003, \mnras, 344, 1000

\bibitem[Calzetti et al.(2000)]{Calzetti2000} Calzetti, D., Armus, L., Bohlin, R.~C., et al.\ 2000, \apj, 533, 682 

\bibitem[Cardamone et al.(2009)]{Cardamone2009} Cardamone, C., Schawinski, K., Sarzi, M., et al.\ 2009, \mnras, 399, 1191 

%\bibitem[{{Cardelli} {et~al.}(1989){Cardelli}, {Clayton}, \&  {Mathis}}]{Cardelli1989} {Cardelli}, J.~A., {Clayton}, G.~C., \& {Mathis}, J.~S. 1989, \apj, 345, 245

\bibitem[Castellano et al.(2014)]{Castellano2014} Castellano, M., Sommariva, V., Fontana, A., et al.\ 2014, \aap, 566, A19 


\bibitem[Cowie et al.(1996)]{Cowie1996} Cowie, L.~L., Songaila, A., Hu, E.~M., \& Cohen, J.~G.\ 1996, \aj, 112, 839 

\bibitem[{{Chabrier}(2003)}]{Chabrier2003} {Chabrier}, G. 2003, \pasp, 115, 763

%\bibitem[Charlot \& Longhetti(2001)]{Charlot2001} Charlot, S., \& Longhetti, M.\ 2001, \mnras, 323, 887 

%\bibitem[Christensen et al.(2012)]{Christensen2012} Christensen, L., Richard, J., Hjorth, J., et al.\ 2012, \mnras, 427, 1953 

%\bibitem[Elbaz et al.(2007)]{Elbaz2007} Elbaz, D., Daddi, E., Le Borgne, D., et al.\ 2007, \aap, 468, 33 

%\bibitem[Erb et al.(2006)]{Erb2006} Erb, D.~K., Shapley, A.~E., Pettini, M., et al.\ 2006, \apj, 644, 813 

\bibitem[Dom{\'{\i}}nguez et al.(2013)]{Dominguez2013} 
Dom{\'{\i}}nguez, A., Siana, B., Henry, A.~L., et al.\ 2013, \apj, 763, 145 

\bibitem[Fosbury et al.(2003)]{Fosbury2003} Fosbury, R.~A.~E., Villar-Mart{\'{\i}}n, M., Humphrey, A., et al.\ 2003, \apj, 596, 797 

%\bibitem[Grogin et al.(2011)]{Grogin2011} Grogin, N.~A., Kocevski, D.~D., Faber, S.~M., et al.\ 2011, \apjs, 197, 35 

%\bibitem[Guaita et al.(2013)]{Guaita2013} Guaita, L., Francke, H., Gawiser, E., et al.\ 2013, \aap, 551, A93

\bibitem[Guseva et al.(2009)]{Guseva2009} Guseva, N.~G., Papaderos, P., Meyer, H.~T., % Izotov, Y.~I., \& Fricke, K.~J.
et al. \ 2009, \aap, 505, 63 

\bibitem[H{\"a}gele et al.(2008)]{Hagele2008} H{\"a}gele, G.~F., D{\'{\i}}az, {\'A}.~I., Terlevich, E., et al. \ 2008, \mnras, 383, 209 

%\bibitem[Hayes et al.(2013)]{Hayes2013} Hayes, M., {\"O}stlin, G., Schaerer, D., et al.\ 2013, \apjl, 765, L27 

%\bibitem[Henry et al.(2013)]{Henry2013b} Henry, A., Scarlata, C., Dom{\'{\i}}nguez, A., et al.\ 2013b, \apjl, 776, L27 


\bibitem[Henry et al.(2013)]{Henry2013} Henry, A., Martin, C.~L., Finlator, K., \& Dressler, A.\ 2013, \apj, 769, 148 

%\bibitem[Hopkins \& Beacom(2006)]{Hopkins2006} Hopkins, A.~M., \& Beacom, J.~F.\ 2006, \apj, 651, 142 

\bibitem[Hoyos et al.(2005)]{Hoyos2005} Hoyos, C., Koo, D.~C., Phillips, et al.\ 2005, \apjl, 635, L21 

\bibitem[Hu et al.(2009)]{Hu2009} Hu, E.~M., Cowie, L.~L., Kakazu, Y., \& Barger, A.~J.\ 2009, \apj, 698, 2014 

%\bibitem[Hunt et al.(2012)]{Hunt2012} Hunt, L., Magrini, L., Galli, D., et al.\ 2012, arXiv:1209.1100 

%\bibitem[Izotov et al.(2006)]{Izotov2006} Izotov, Y.~I., Papaderos, P., Guseva, N.~G., Fricke, K.~J., \& Thuan, T.~X.\ 2006, \aap, 454, 137 

\bibitem[Jaskot \& Oey(2013)]{Jaskot2013} Jaskot, A.~E., \& Oey, M.~S.\ 2013, \apj, 766, 91 

\bibitem[Juneau et al.(2011)]{Juneau2011} Juneau, S., Dickinson, M., Alexander, D.~M., \& Salim, S.\ 2011, \apj, 736, 104 

%\bibitem[Kakazu et al.(2007)]{Kakazu2007} Kakazu, Y., Cowie, L.~L., \& Hu, E.~M.\ 2007, \apj, 668, 853 

%\bibitem[{{Kauffmann} {et~al.}(2003){Kauffmann}, {Heckman}, {Tremonti}, {Brinchmann}, {Charlot}, {White}, {Ridgway}, {Brinkmann}, {Fukugita}, {Hall},  {Ivezi{\'c}}, {Richards}, \& {Schneider}}]{Kauffmann2003} {Kauffmann}, G., {Heckman}, T.~M., {Tremonti}, C., {et~al.} 2003, \mnras, 346, 1055

\bibitem[Kennicutt(1998)]{Kennicutt1998} Kennicutt, R.~C., Jr.\ 1998, \apj, 498, 541 

%\bibitem[Kewley et al.(2001)]{Kewley2001} Kewley, L.~J., Dopita, M.~A., Sutherland, R.~S., Heisler, C.~A., \& Trevena, J.\ 2001, \apj, 556, 121 

%\bibitem[{{Kewley} \& {Dopita}(2002)}]{KewleyDopita2002} {Kewley}, L.~J. \& {Dopita}, M.~A. 2002, \apjs, 142, 35

%\bibitem[Kewley et al.(2006)]{Kewley2006} Kewley, L.~J., Groves, B., Kauffmann, G., \& Heckman, T.\ 2006, \mnras, 372, 961 

\bibitem[Kewley et al.(2007)]{Kewley2007} Kewley, L.~J., Brown, W.~R., Geller, M.~J., Kenyon, S.~J., \& Kurtz, M.~J.\ 2007, \aj, 133, 882 

%\bibitem[Kewley \& Ellison(2008)]{Kewley2008} Kewley, L.~J., \& Ellison, S.~L.\ 2008, \apj, 681, 1183 

%\bibitem[Kniazev et al.(2003)]{Kniazev2003} Kniazev, A.~Y., Grebel, E.~K., Hao, L., et al.\ 2003, \apjl, 593, L73 

\bibitem[Kniazev et al.(2004)]{Kniazev2004} Kniazev, A.~Y., Pustilnik, S.~A., Grebel, E.~K., Lee, H., \& Pramskij, A.~G.\ 2004, \apjs, 153, 429 

%\bibitem[Kobulnicky et al.(2003)]{Kobulnicky2003} Kobulnicky, H.~A., Willmer, C.~N.~A., Phillips, A.~C., et al.\ 2003, \apj, 599, 1006 

%\bibitem[Lara-L{\'o}pez et al.(2010)]{LaraLopez2010} Lara-L{\'o}pez, M.~A., Cepa, J., Bongiovanni, A., et al.\ 2010, \aap, 521, L53
 
%\bibitem[Koekemoer et al.(2011)]{Koekemoer2011} Koekemoer, A.~M., Faber, S.~M., Ferguson, H.~C., et al.\ 2011, \apjs, 197, 36 

%\bibitem[Kobulnicky et al.(2003)]{Kobulnicky2003} Kobulnicky, H.~A., Willmer, C.~N.~A., Phillips, A.~C., et al.\ 2003, \apj, 599, 1006 

%\bibitem[Kunth \& \"Ostlin(2000)]{KunthOstlin2000} Kunth, D., {\"O}stlin, G.\ 2000, \aapr, 10, 1 

%\bibitem[Lamareille et al.(2004)]{Lamareille2004} Lamareille, F., Mouhcine, M., Contini, T., Lewis, I., \& Maddox, S. 2004, MNRAS, 350, 396
\bibitem[Lamareille et al.(2006)]{Lamareille2006} Lamareille, F., Contini, T., Brinchmann, J., et al.\ 2006b, \aap, 448, 907

\bibitem[Lamareille et al.(2009)]{Lamareille2009} Lamareille, F., Brinchmann, J., Contini, T., et al.\ 2009, \aap, 495, 53 

\bibitem[Le F{\`e}vre et al.(2003)]{LeFevre2003} Le F{\`e}vre, O., Saisse, M., Mancini, D., et al.\ 2003, \procspie, 4841, 1670 

\bibitem[Le Fevre et al.(2014)]{LeFevre2014} Le Fevre, O., Tasca, 
L.~A.~M., Cassata, P., et al.\ 2014, arXiv:1403.3938 


\bibitem[Ly et al.(2014)]{Ly2014} Ly, C., Malkan, M.~A., Nagao, T., et al.\ 2014, \apj, 780, 122 

%\bibitem[Lequeux et al.(1979)]{Lequeux1979} Lequeux, J., Peimbert, M., Rayo, J.~F., Serrano, A., \& Torres-Peimbert, S.\ 1979, \aap, 80, 155 

\bibitem[Maiolino et al.(2008)]{Maiolino2008} Maiolino, R., Nagao, T., Grazian, A., et al.\ 2008, \aap, 488, 463 

%\bibitem[Mannucci et al.(2009)]{Mannucci2009} Mannucci, F., Cresci, G., Maiolino, R., et al.\ 2009, \mnras, 398, 1915 

\bibitem[Mannucci et al.(2010)]{Mannucci2010} Mannucci, F., Cresci, G., Maiolino, R., Marconi, A., \& Gnerucci, A.\ 2010, \mnras, 408, 2115 

\bibitem[Mannucci et al.(2011)]{Mannucci2011} Mannucci, F., Salvaterra, R., \& Campisi, M.~A.\ 2011, \mnras, 414, 1263 

%\bibitem[Maraston(2005)]{Maraston2005} Maraston, C.\ 2005, \mnras, 362, 799 

%\bibitem[Maseda et al.(2013)]{Maseda2013} Maseda, M.~V., van der Wel, A., da Cunha, E., et al.\ 2013, \apjl, 778, L22 

\bibitem[Maseda et al.(2014)]{Maseda2014} Maseda, M.~V., van der Wel, A., Rix, H.-W., et al.\ 2014, arXiv:1406.3351 

%\bibitem[Marocco et al.(2011)]{Marocco2011} Marocco, J., Hache, E., \& Lamareille, F.\ 2011, \aap, 531, A71 

\bibitem[McGaugh(1991)]{McGaugh1991} McGaugh, S.~S.\ 1991, \apj, 380, 140 

%\bibitem[Meurer et al.(1999)]{Meurer1999} Meurer, G.~R., Heckman, T.~M., \& Calzetti, D.\ 1999, \apj, 521, 64 

\bibitem[Morales-Luis et al.(2011)]{Morales-Luis2011} Morales-Luis, A.~B., S{\'a}nchez Almeida, J., Aguerri, J.~A.~L., \& Mu{\~n}oz-Tu{\~n}{\'o}n, C.\ 2011, \apj, 743, 77 

%\bibitem[Nagao et al.(2006)]{Nagao2006} Nagao, T., Maiolino, R., \& Marconi, A.\ 2006, \aap, 459, 85 

\bibitem[Nakajima \& Ouchi(2014)]{Nakajima2013} Nakajima, K., \& Ouchi, M.\ 2014, \mnras, 442, 900 

%\bibitem[Noeske et al.(2007)]{Noeske2007} Noeske, K.~G., Weiner, B.~J., Faber, S.~M., et al.\ 2007, \apjl, 660, L43 

%\bibitem[Pagel et al.(1979)]{Pagel1979} Pagel, B.~E.~J., Edmunds, M.~G., Blackwell, D.~E., Chun, M.~S., \& Smith, G.\ 1979, \mnras, 189, 95 

%\bibitem[Papaderos et al.(2006)]{Papaderos2006} Papaderos, P., Guseva, N.~G., Izotov, Y.~I., et al.\ 2006, \aap, 457, 45 

\bibitem[Papaderos et al.(2008)]{Papaderos2008} Papaderos, P., Guseva, N.~G., Izotov, Y.~I., \& Fricke, K.~J.\ 2008, \aap, 491, 113 

\bibitem[P\'erez-Montero \& D\'iaz(2003)]{Perez-Montero2003} P\'erez-Montero, E. \& D\'iaz, A.I. 2003, MNRAS, 346, 105.

\bibitem[{{P{\'e}rez-Montero} \& {Contini}(2009)}]{Perez-Montero2009}{P{\'e}rez-Montero}, E. \& {Contini}, T. 2009, \mnras, 398, 949

%\bibitem[{{P{\'e}rez-Montero} \& {D{\'{\i}}az}(2005)}]{Perez-Montero2005}{P{\'e}rez-Montero}, E. \& {D{\'{\i}}az}, A.~I. 2005, \mnras, 361, 1063

\bibitem[P{\'e}rez-Montero et al.(2013)]{Perez-Montero2013} P{\'e}rez-Montero, E., Contini, T., Lamareille, F., et al.\ 2013, \aap, 549, A25 

%\bibitem[Pettini \& Pagel(2004)]{PettiniPagel2004} Pettini, M., \& Pagel, B.~E.~J.\ 2004, \mnras, 348, L59

\bibitem[P{\'e}rez-Gonz{\'a}lez et al.(2008)]{Perez-Gonzalez2008} P{\'e}rez-Gonz{\'a}lez, P.~G., Rieke, G.~H., Villar, V., et al.\ 2008, \apj, 675, 234 

%\bibitem[Pilyugin \& Thuan(2005)]{Pilyugin2005} Pilyugin, L.~S., \& Thuan, T.~X.\ 2005, \apj, 631, 231 

%\bibitem[Richard et al.(2011)]{Richard2011} Richard, J., Jones, T., Ellis, R., et al.\ 2011, \mnras, 413, 643 

%\bibitem[Salim et al.(2007)]{Salim2007} Salim, S., Rich, R.~M., Charlot, S., et al.\ 2007, \apjs, 173, 267 
\bibitem[S{\'a}nchez Almeida et al.(2014)]{SanchezAlmeida2014} S{\'a}nchez 
Almeida, J., Morales-Luis, A.~B., et al.\ 2014, \apj, 783, 45 

%\bibitem[Santini et al.(2012)]{Santini2012} Santini, P., Fontana, A., Grazian, A., et al.\ 2012, \aap, 538, A33 



%\bibitem[Schiminovich et al.(2007)]{Schiminovich2007} Schiminovich, D., Wyder, T.~K., Martin, D.~C., et al.\ 2007, \apjs, 173, 315 
%\bibitem[Seifert et al.(2003)]{Seifert2003} Seifert, W. et al. 2003, Proc. SPIE, 4841, 962

%\bibitem[Storey \& Hummer(1995)]{StoreyHummer1995} Storey, P.~J., \& Hummer, D.~G.\ 1995, \mnras, 272, 41 

%\bibitem[Straughn et al.(2006)]{Straughn2006} Straughn, A.~N., Cohen, S.~H., Ryan, R.~E., et al.\ 2006, \apj, 639, 724 

%\bibitem[Terlevich et al.(1991)]{Terlevich1991} Terlevich, R., Melnick, J., Masegosa, J., Moles, M., \& Copetti, M.~V.~F.\ 1991, \aaps, 91, 285 

%\bibitem[{{Tremonti} {et~al.}(2004){Tremonti}, {Heckman}, {Kauffmann},  {Brinchmann}, {Charlot}, {White}, {Seibert}, {Peng}, {Schlegel}, {Uomoto},
%  {Fukugita}, \& {Brinkmann}}]{Tremonti2004} {Tremonti}, C.~A., {Heckman}, T.~M., {Kauffmann}, G., {et~al.} 2004, \apj, 613, 898
\bibitem[Tasca et al.(2009)]{Tasca2009} Tasca, L.~A.~M., Kneib, J.-P., Iovino, A., et al.\ 2009, \aap, 503, 379 

\bibitem[Troncoso et al.(2014)]{Troncoso2014} Troncoso, P., Maiolino, R., Sommariva, V., et al.\ 2014, \aap, 563, A58 

\bibitem[van der Wel et al.(2011)]{vdWel2011} van der Wel, A., Straughn, A.~N., Rix, H.-W., et al.\ 2011, \apj, 742, 111 

%\bibitem[van der Wel et al.(2013)]{vanderWel2013} van der Wel, A., van de Ven, G., Maseda, M., et al.\ 2013, arXiv:1309.2826 

%\bibitem[Veilleux \& Osterbrock(1987)]{Veilleux1987} Veilleux, S., \& Osterbrock, D.~E.\ 1987, \apjs, 63, 295 

%\bibitem[Whitaker et al.(2011)]{Whitaker2011} Whitaker, K.~E., Labb{\'e}, I., van Dokkum, P.~G., et al.\ 2011, \apj, 735, 86 

\bibitem[Whitaker et al.(2012)]{Whitaker2012} Whitaker, K.~E., van Dokkum, P.~G., Brammer, G., \& Franx, M.\ 2012, \apjl, 754, L29 

% \bibitem[Xia et al.(2012)]{Xia2012} Xia, L., Malhotra, S., Rhoads, J., et al.\ 2012, \aj, 144, 28 

\bibitem[Yan et al.(2011)]{Yan2011} Yan, R., Ho, L.~C., Newman, J.~A., et al.\ 2011, \apj, 728, 38 

\bibitem[Zahid et al.(2011)]{Zahid2011} Zahid, H.~J., Kewley, L.~J., \& Bresolin, F.\ 2011, \apj, 730, 137 

%\bibitem[Zahid et al.(2012)]{Zahid2012} Zahid, H.~J., Bresolin, F., Kewley, L.~J., et al. \ 2012, \apj, 750, 120 






\end{thebibliography}
%\end{document}

%\clearpage
%\newpage
%\Online

%\begin{appendix}
%\section{Atlas of galaxy spectra}

%-------------------------------------------------------------
 \onlfig{1}{
%\Online
 \begin{figure*}[t!]
   \includegraphics[angle=0,width=8.5cm]{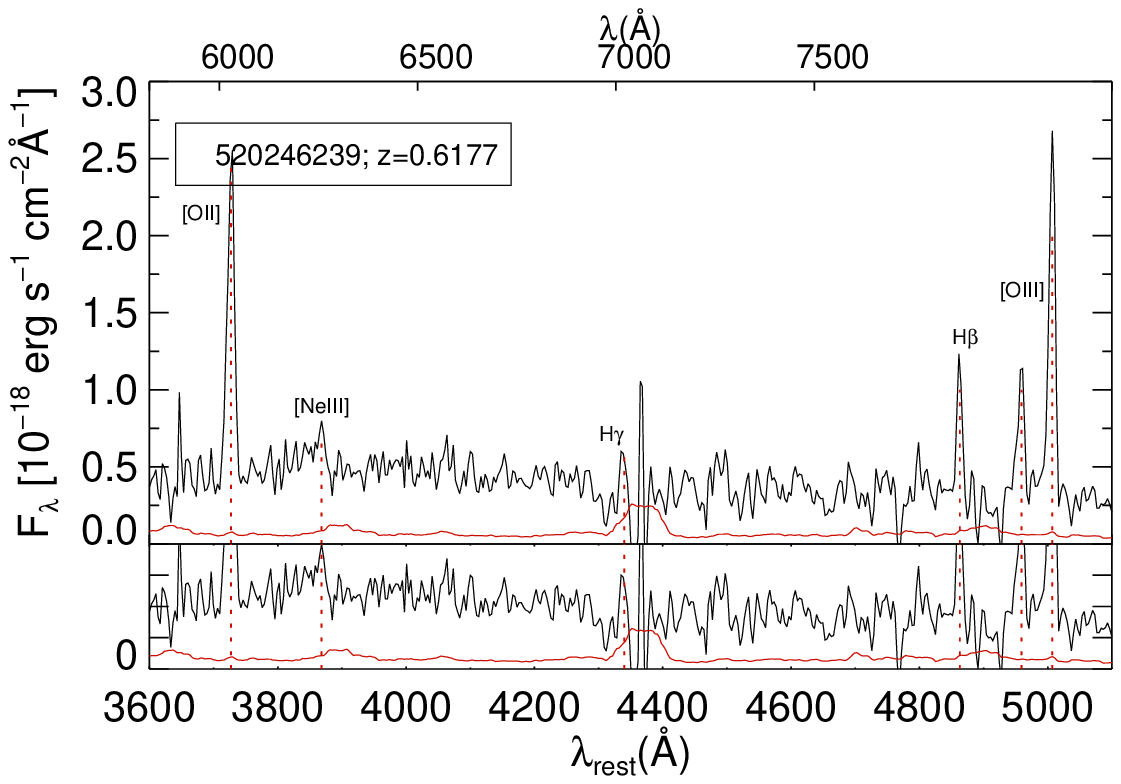} \hspace*{0.5cm}
   \includegraphics[angle=0,width=8.5cm]{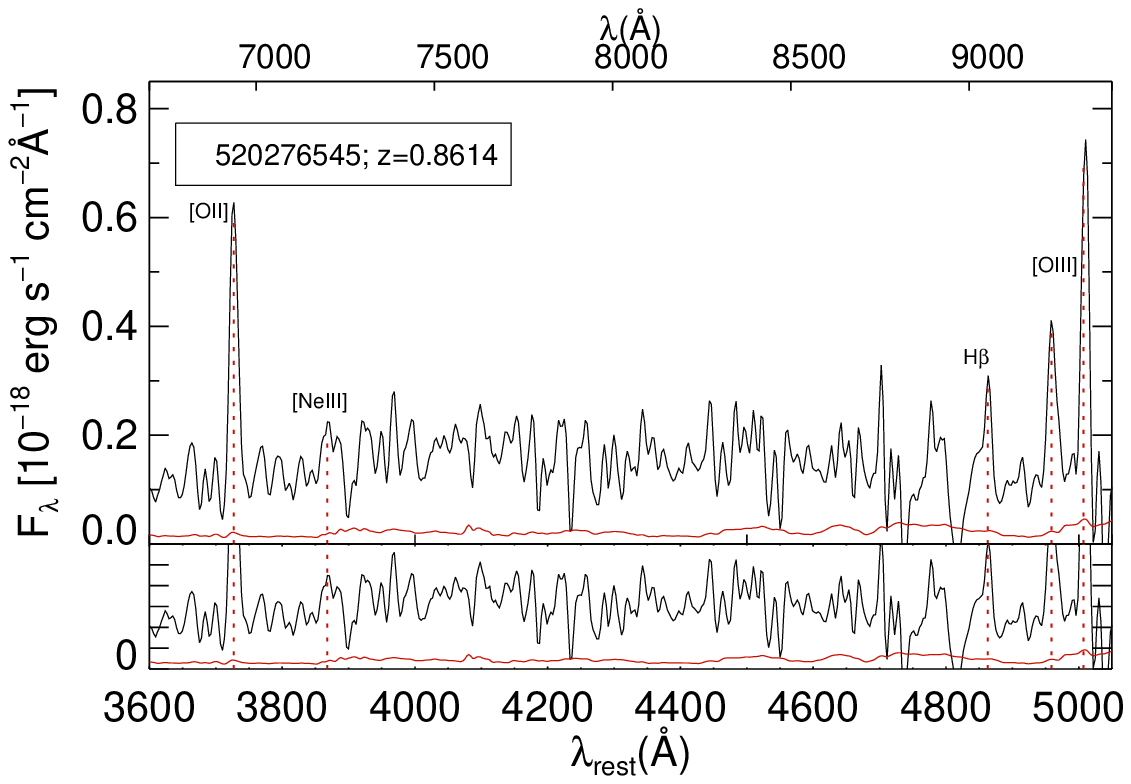} \\[0.5cm]
   \includegraphics[angle=0,width=8.5cm]{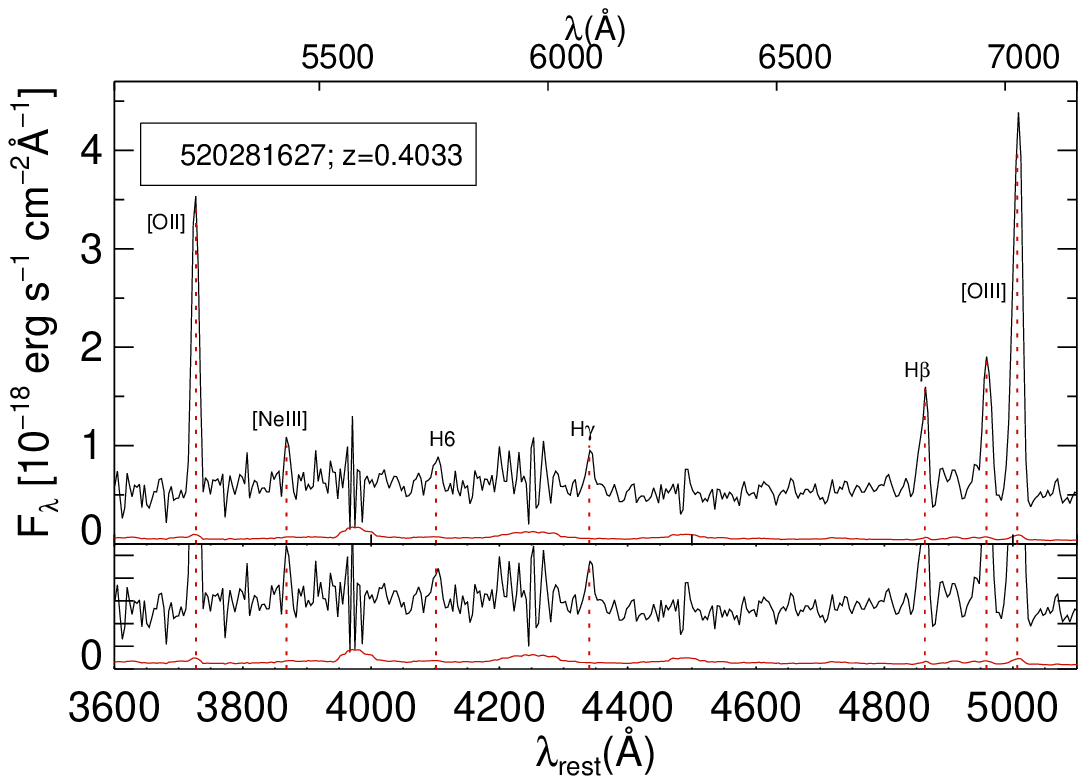} \hspace*{0.5cm}
   \includegraphics[angle=0,width=8.5cm]{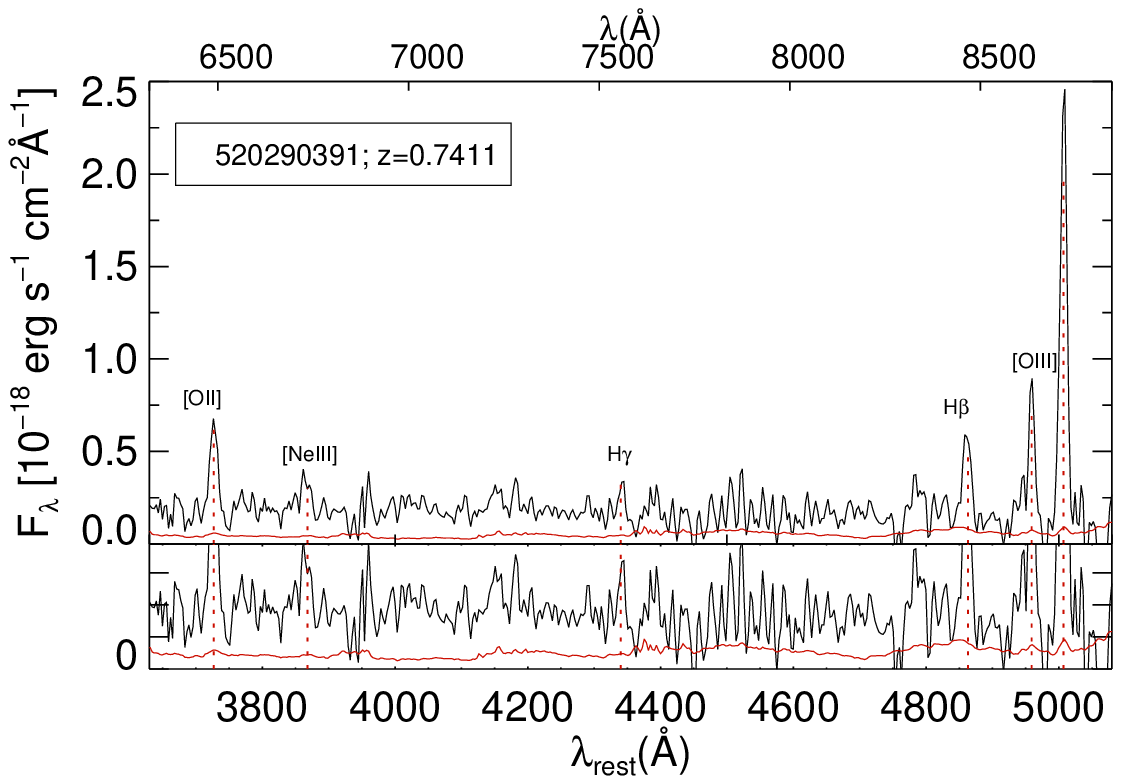} \\[0.5cm]
   \includegraphics[angle=0,width=8.5cm]{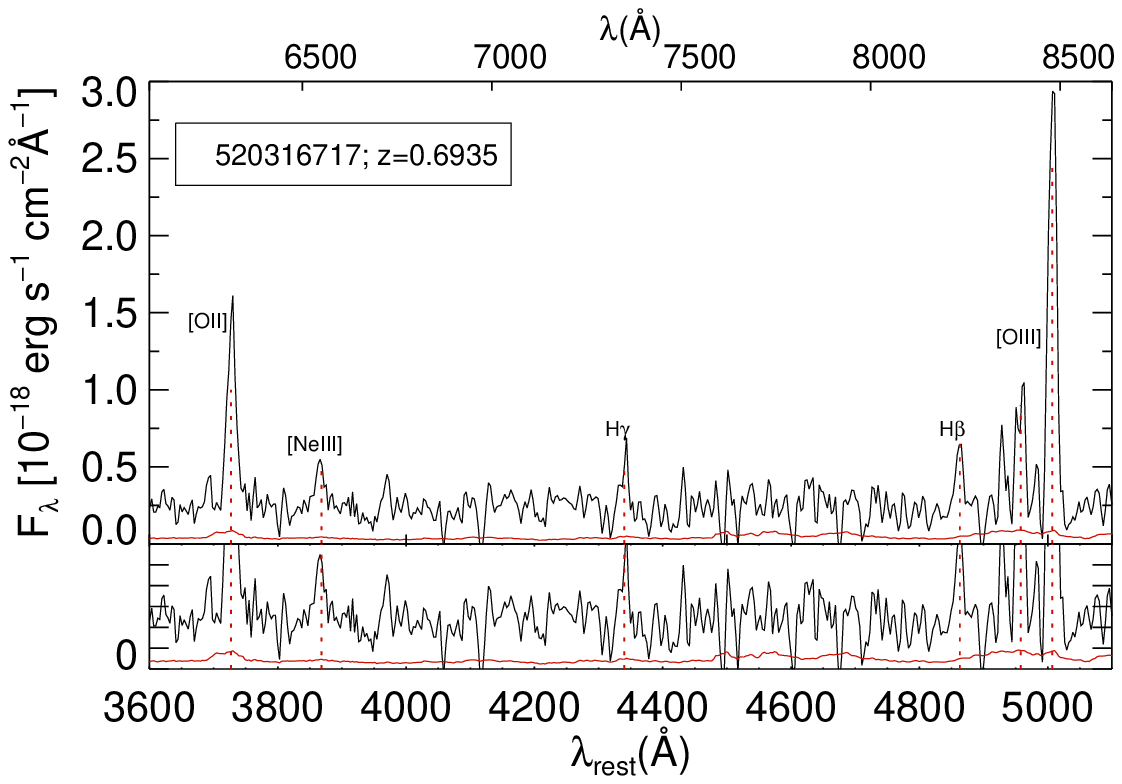} \hspace*{0.5cm}
   \includegraphics[angle=0,width=8.5cm]{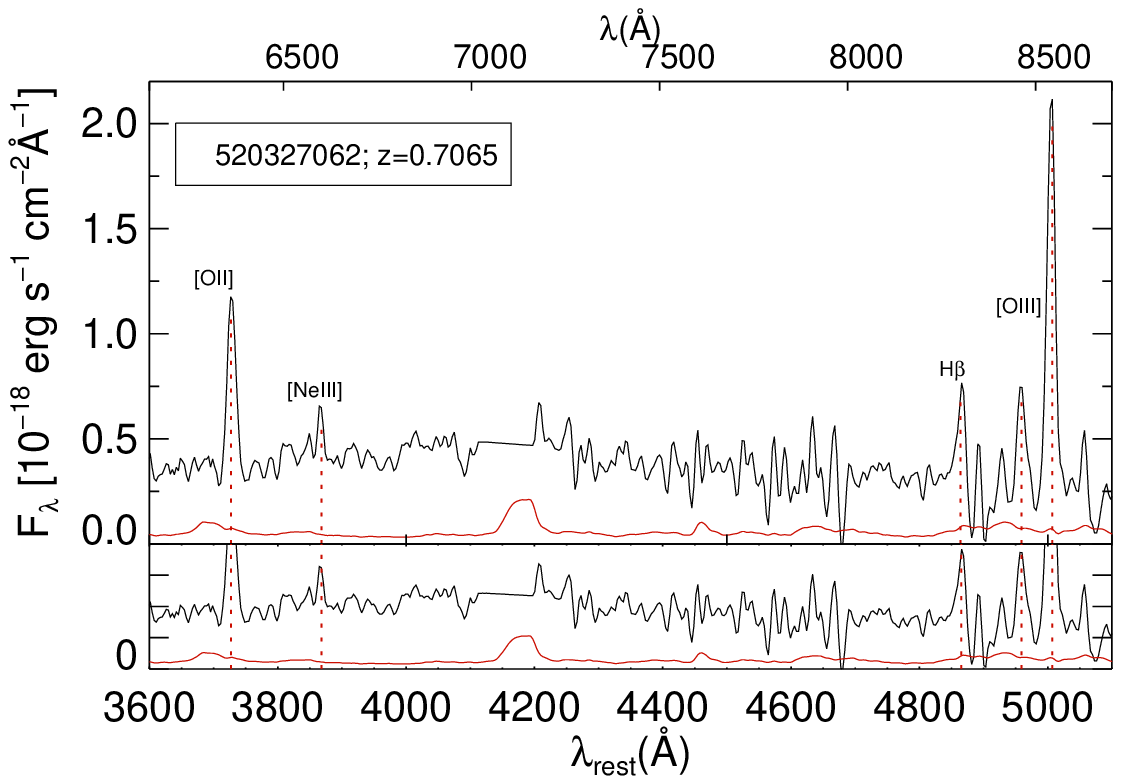} \\[0.5cm]
   \includegraphics[angle=0,width=8.5cm]{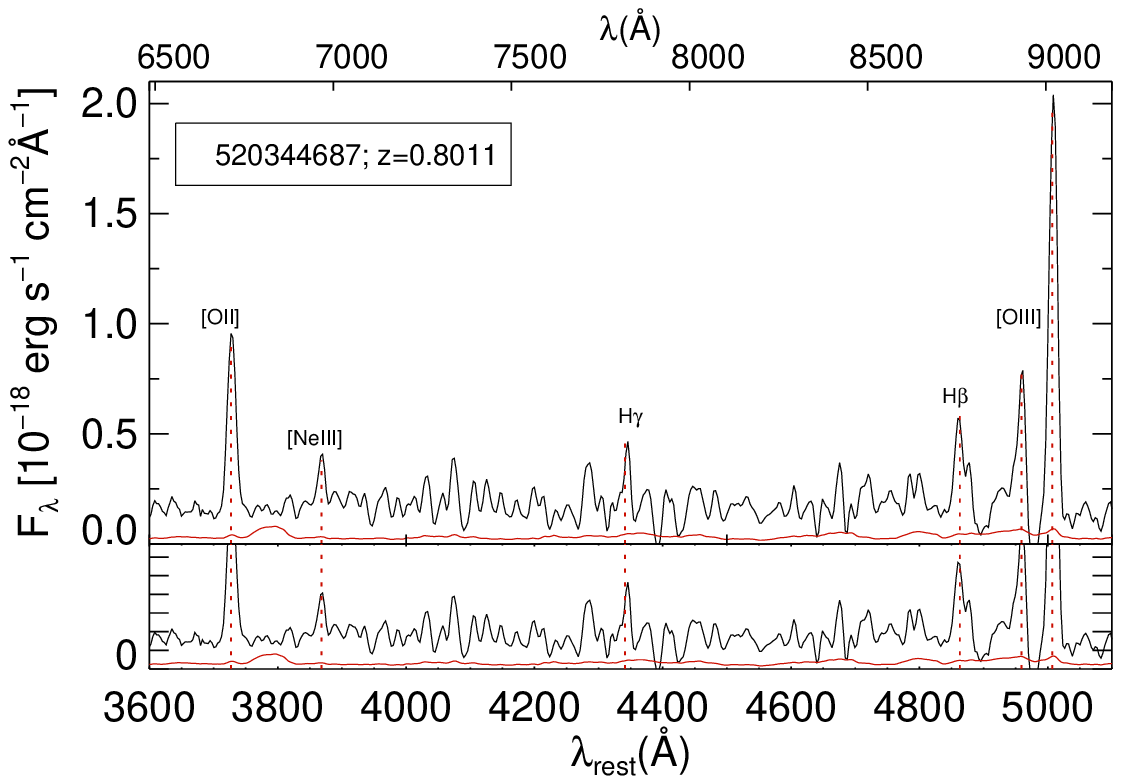} \hspace*{0.5cm}
   \includegraphics[angle=0,width=8.5cm]{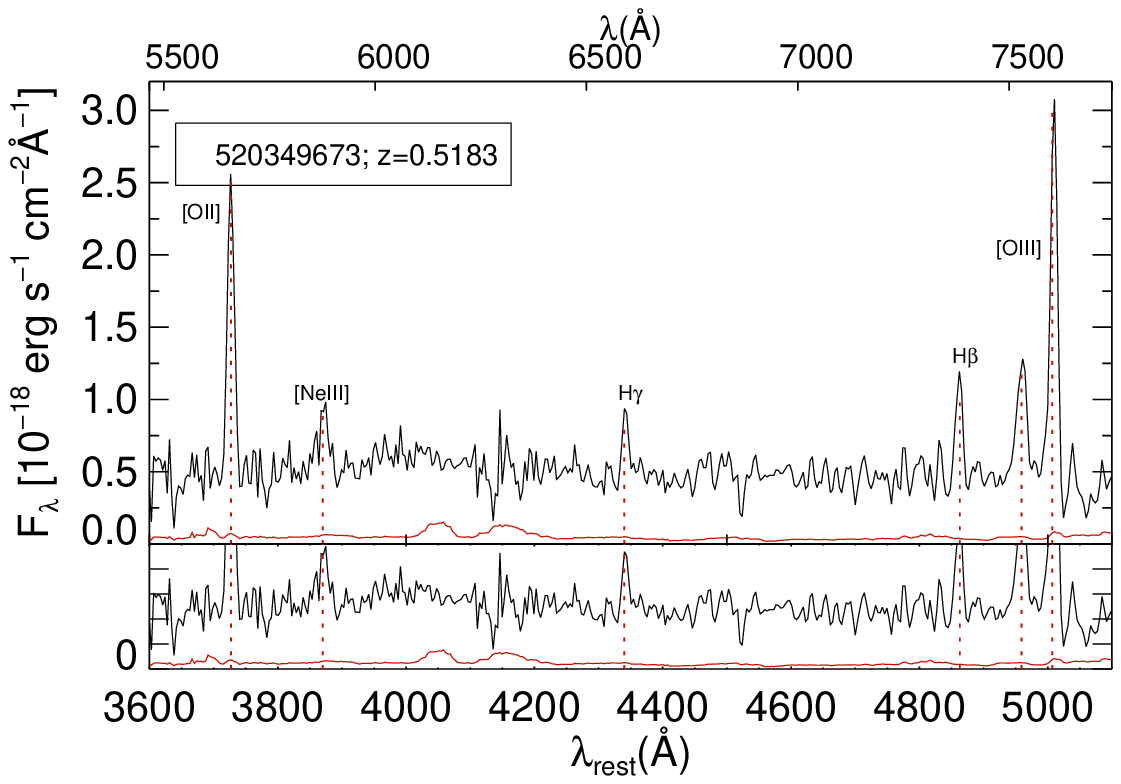} \\[0.5cm]
     \label{atlas}\caption{{VUDS spectra for the sample galaxies. 
     A close-up version is shown in the bottom panel. 
  The galaxy ID, redshift, and the main detected nebular emission lines are labeled. 
  The 1$\sigma$ noise spectrum is shown in red. }}
   \end{figure*}
  
   \begin{figure*}[t!] 
   \ContinuedFloat
   \includegraphics[angle=0,width=8.5cm]{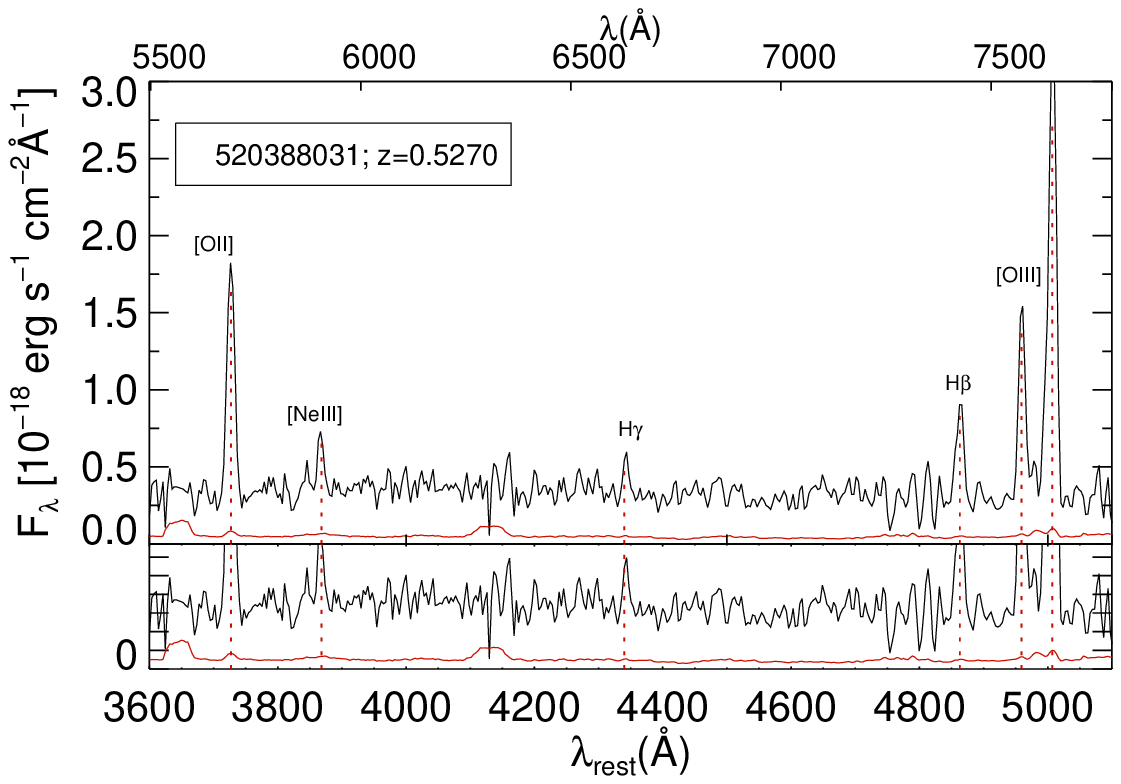} \hspace*{0.5cm}
   \includegraphics[angle=0,width=8.5cm]{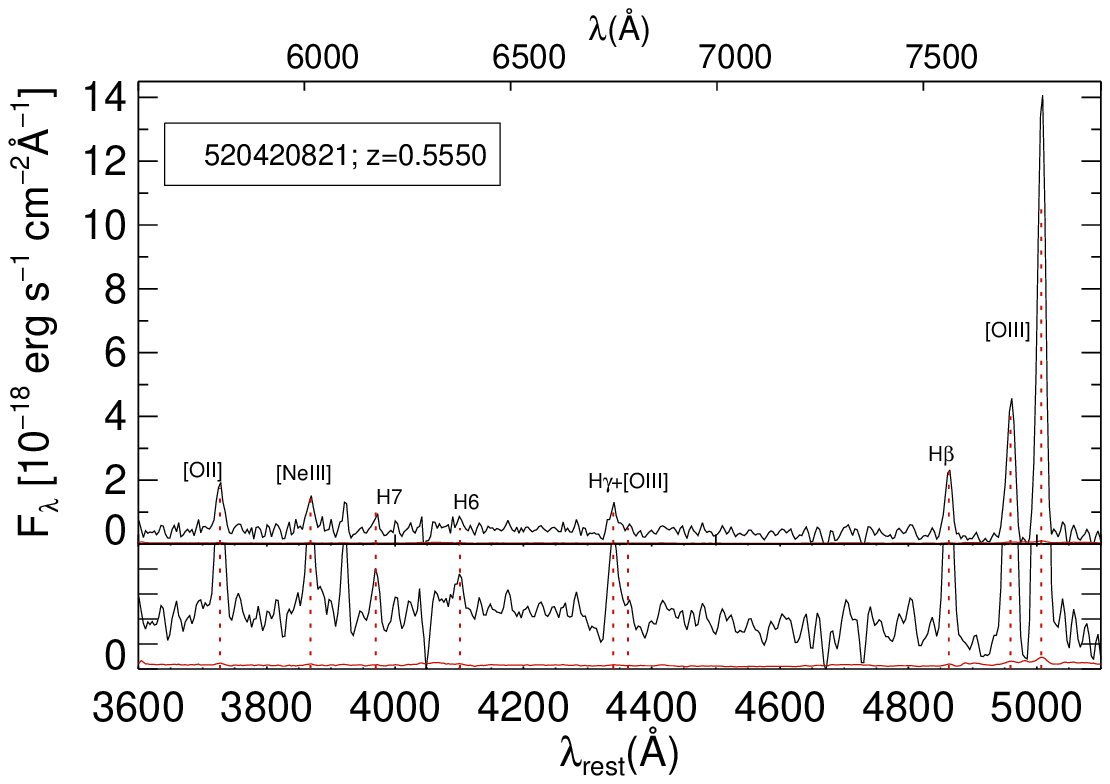} \\[0.5cm]
   \includegraphics[angle=0,width=8.5cm]{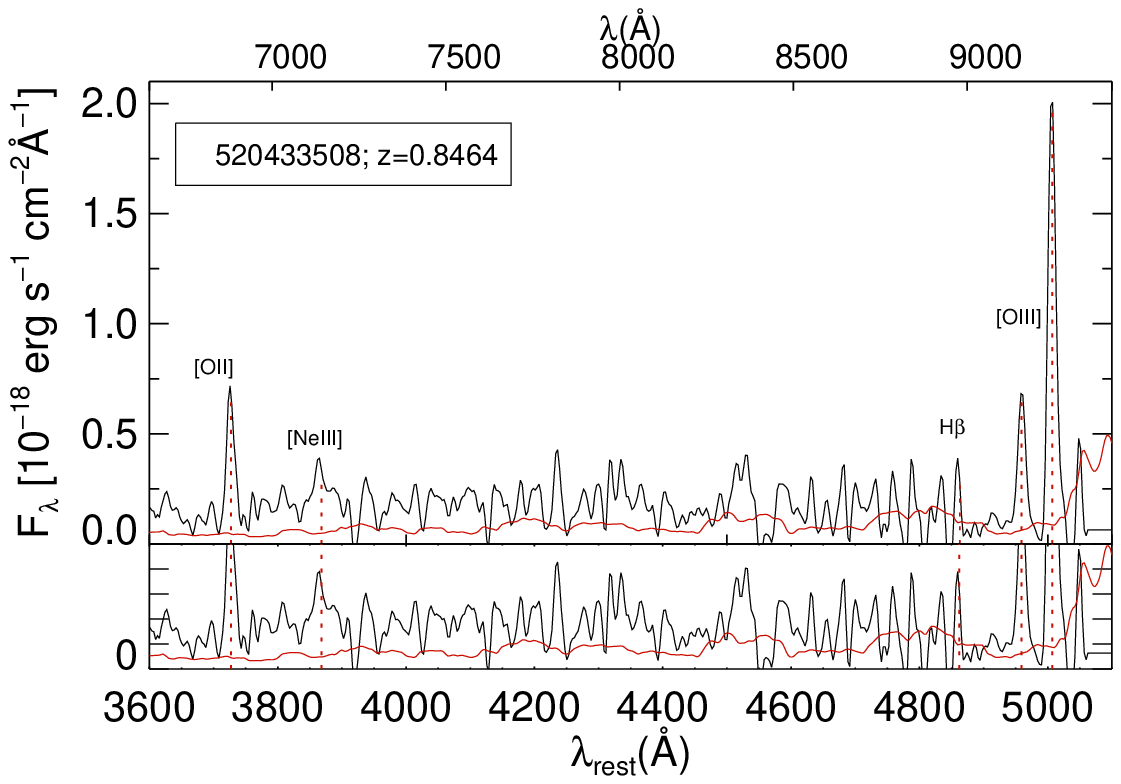} \hspace*{0.5cm}
   \includegraphics[angle=0,width=8.5cm]{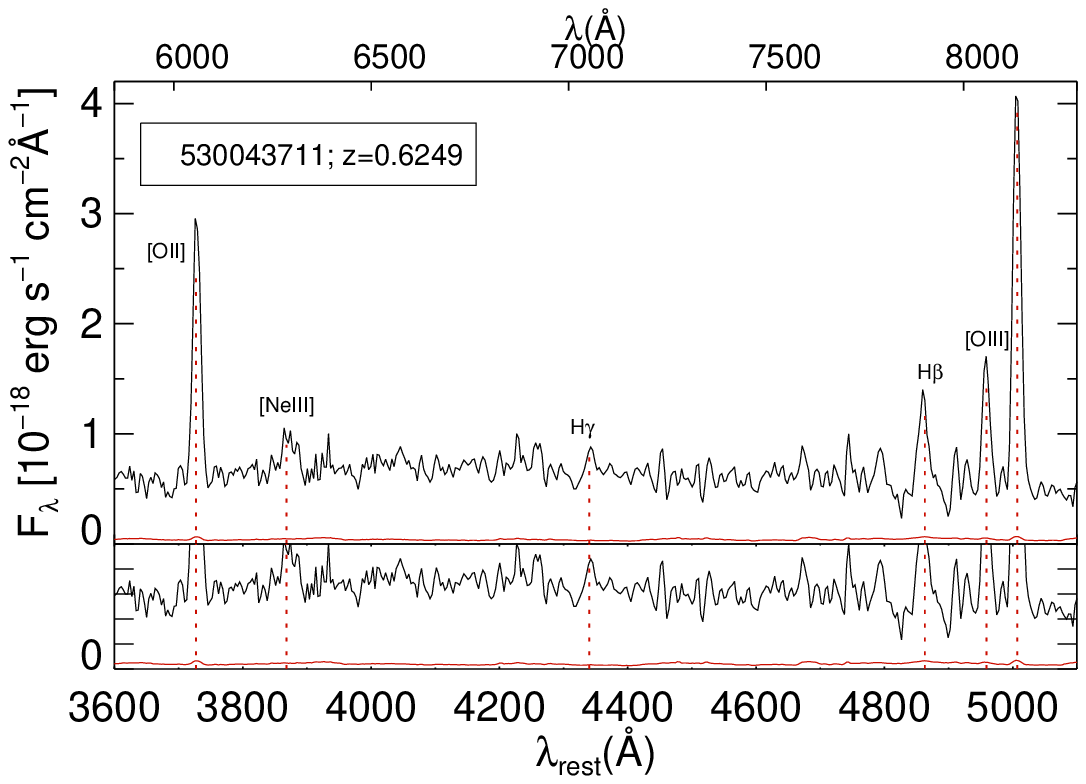} \\[0.5cm]
   \includegraphics[angle=0,width=8.5cm]{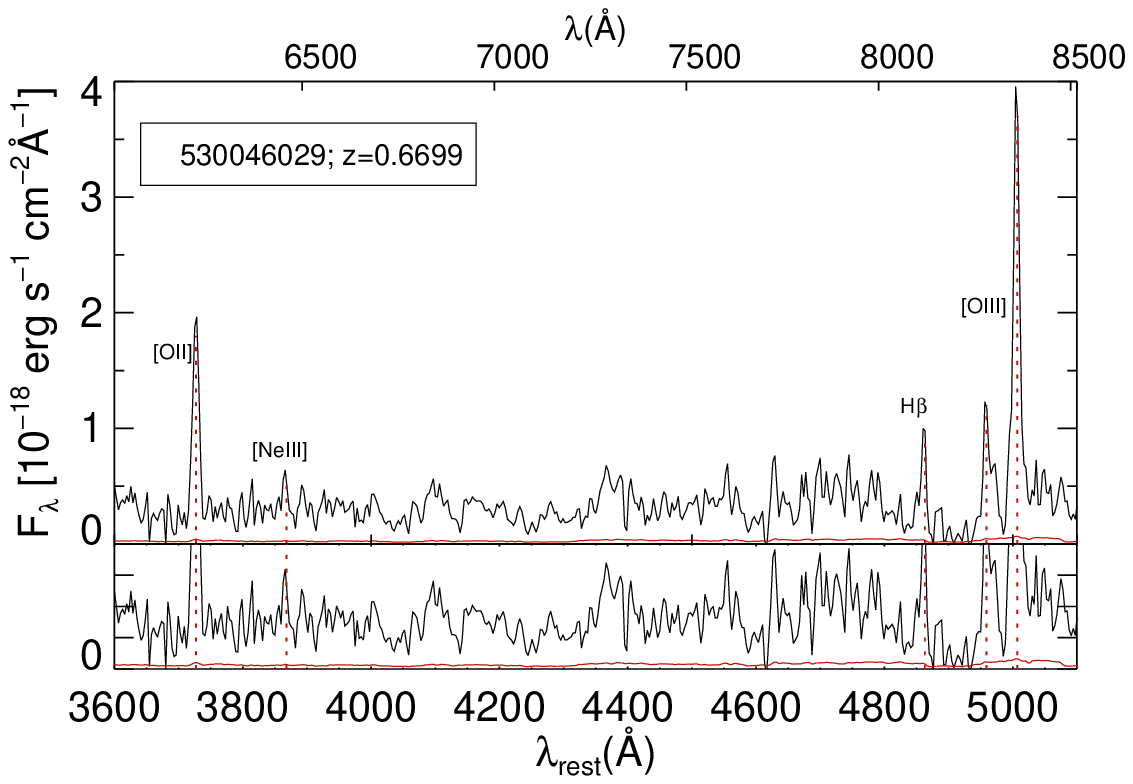} \hspace*{0.5cm}
   \includegraphics[angle=0,width=8.5cm]{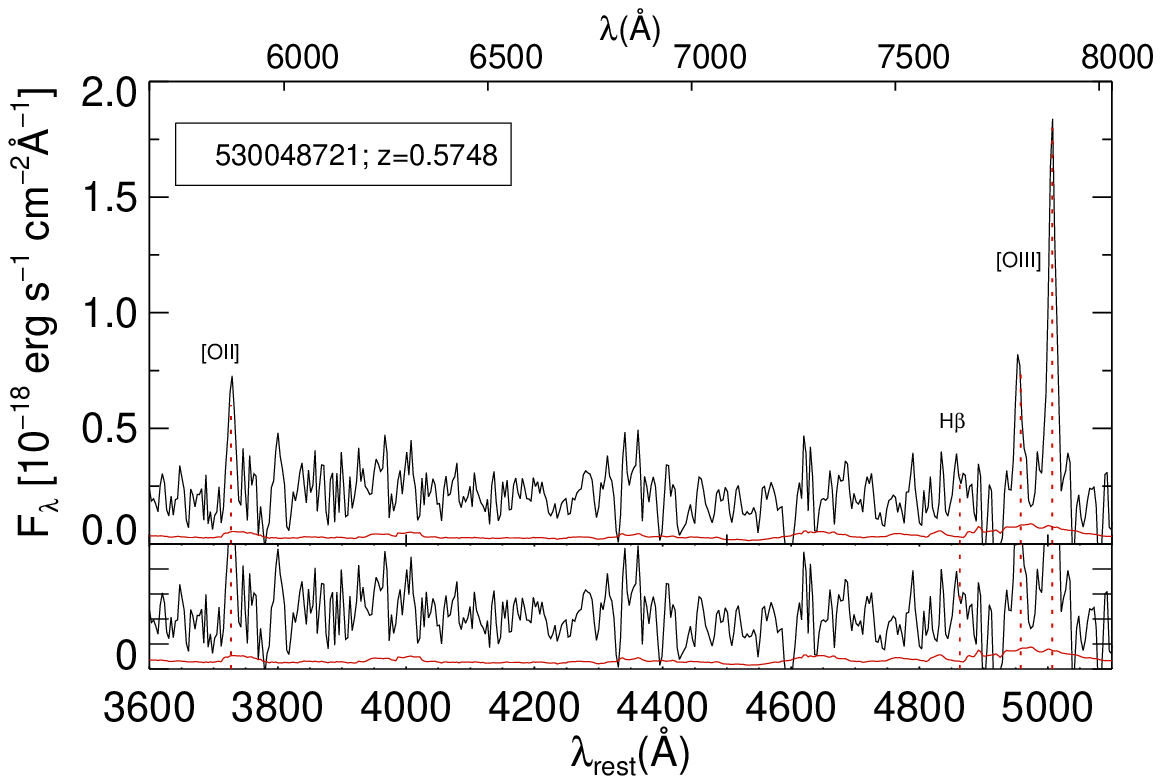} \\[0.5cm]
   \includegraphics[angle=0,width=8.5cm]{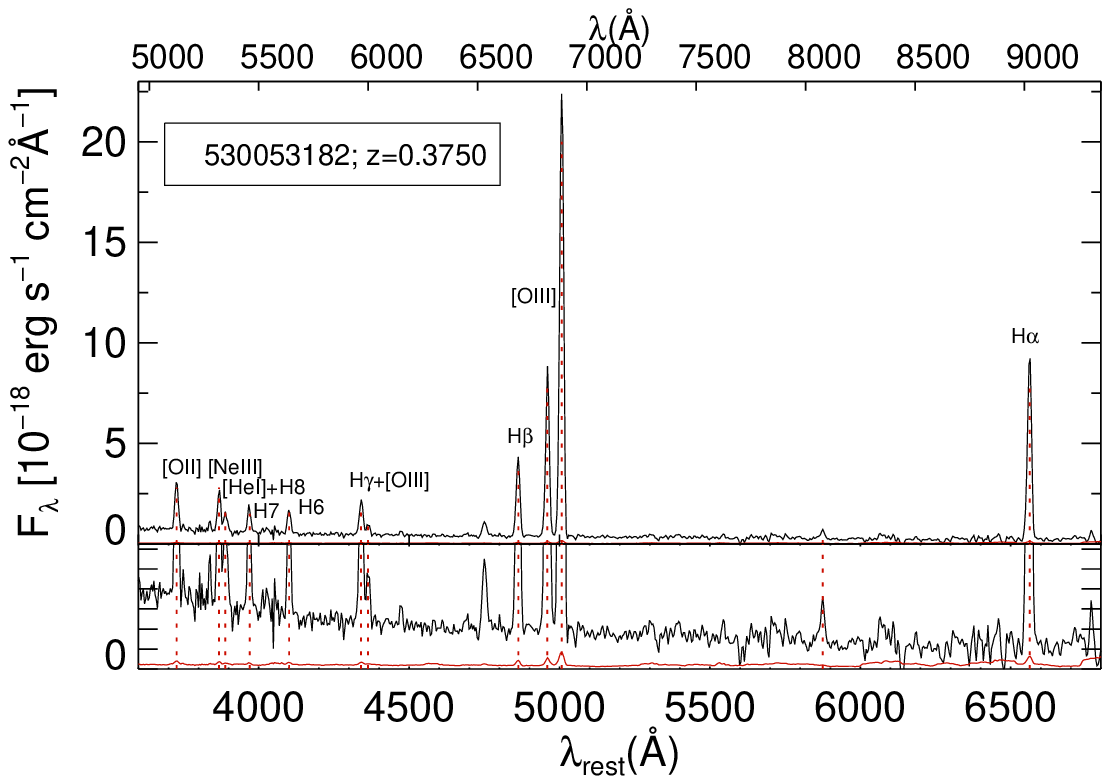} \hspace*{0.5cm}
   \includegraphics[angle=0,width=8.5cm]{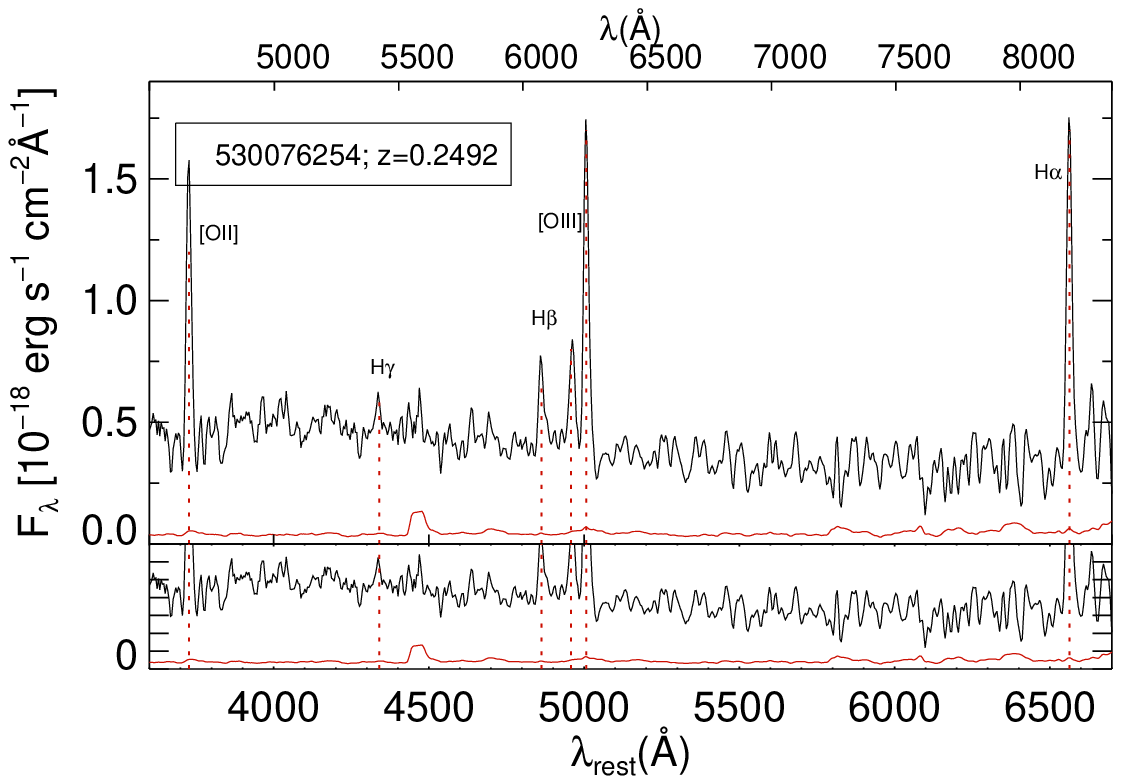} \\[0.5cm]
\caption[]{\label{atlas} VUDS spectra for the sample galaxies. Continued}
   \end{figure*}

   \begin{figure*}[t!] 
   \ContinuedFloat
   \includegraphics[angle=0,width=8.5cm]{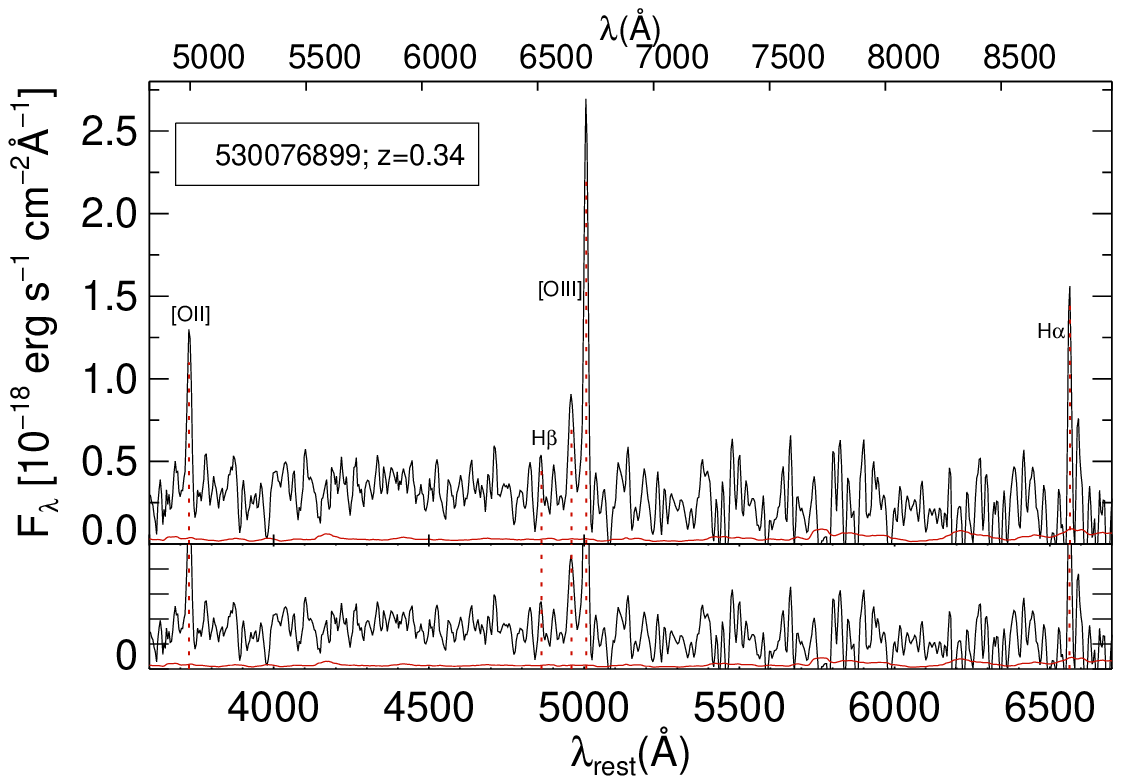} \hspace*{0.5cm}
   \includegraphics[angle=0,width=8.5cm]{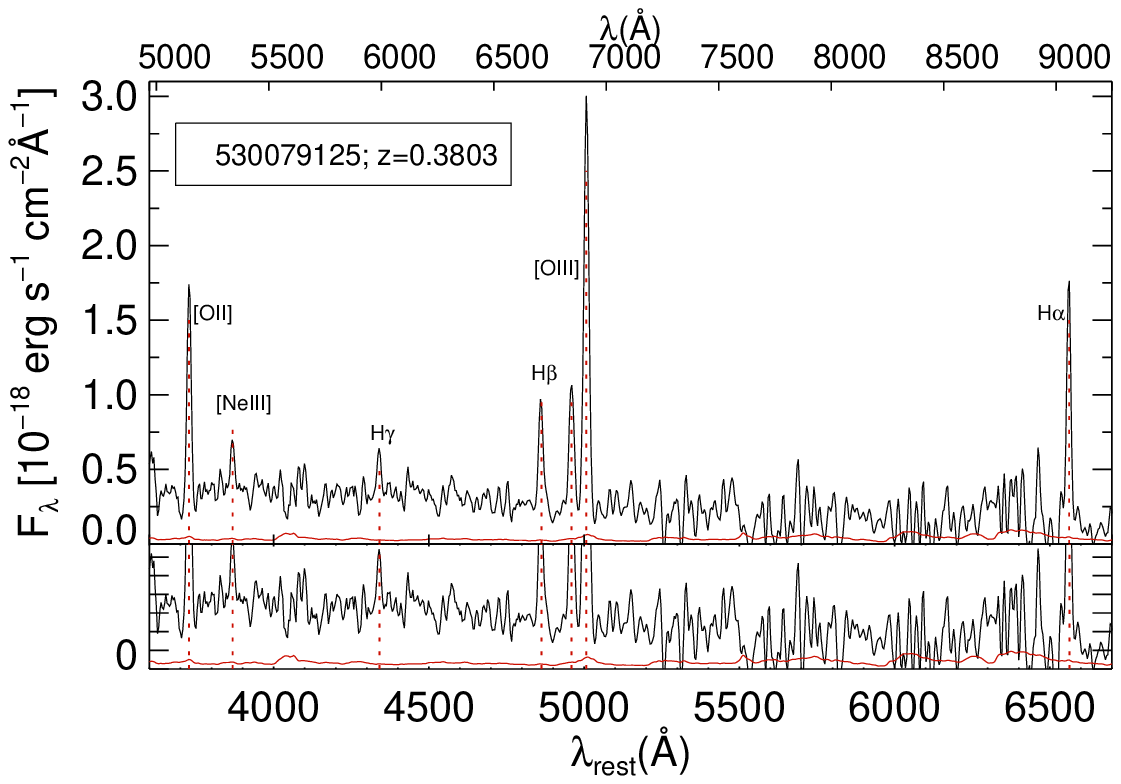} \\[0.5cm]
   \includegraphics[angle=0,width=8.5cm]{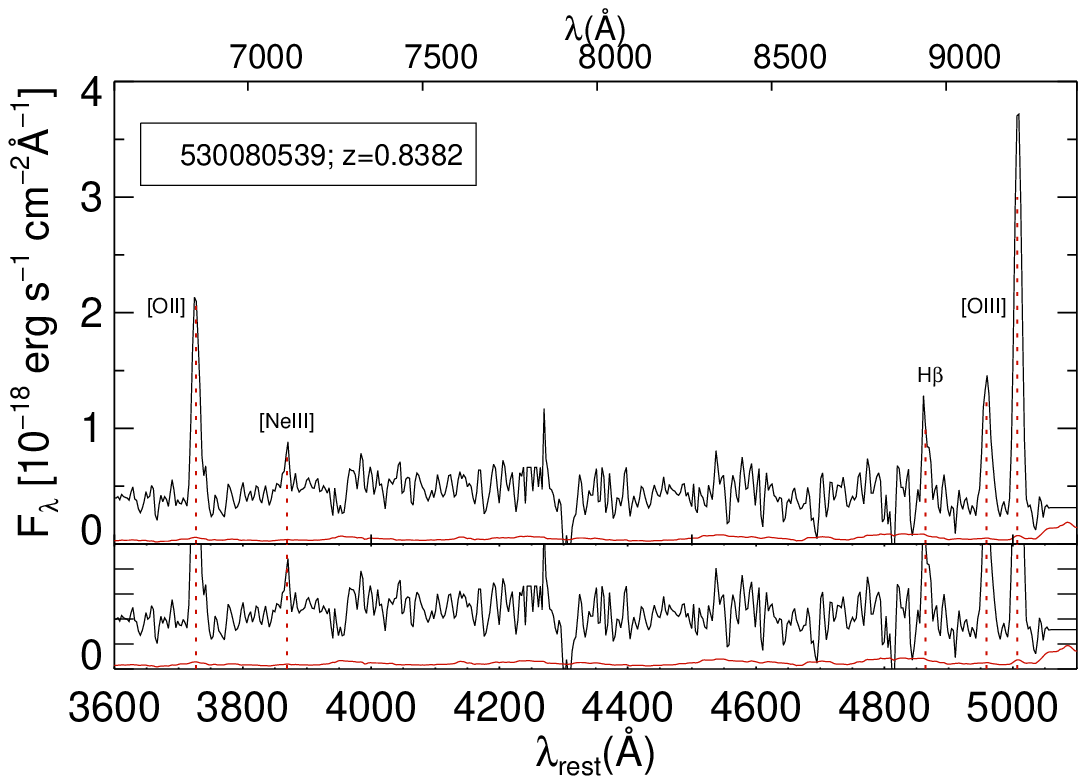} \hspace*{0.5cm}
   \includegraphics[angle=0,width=8.5cm]{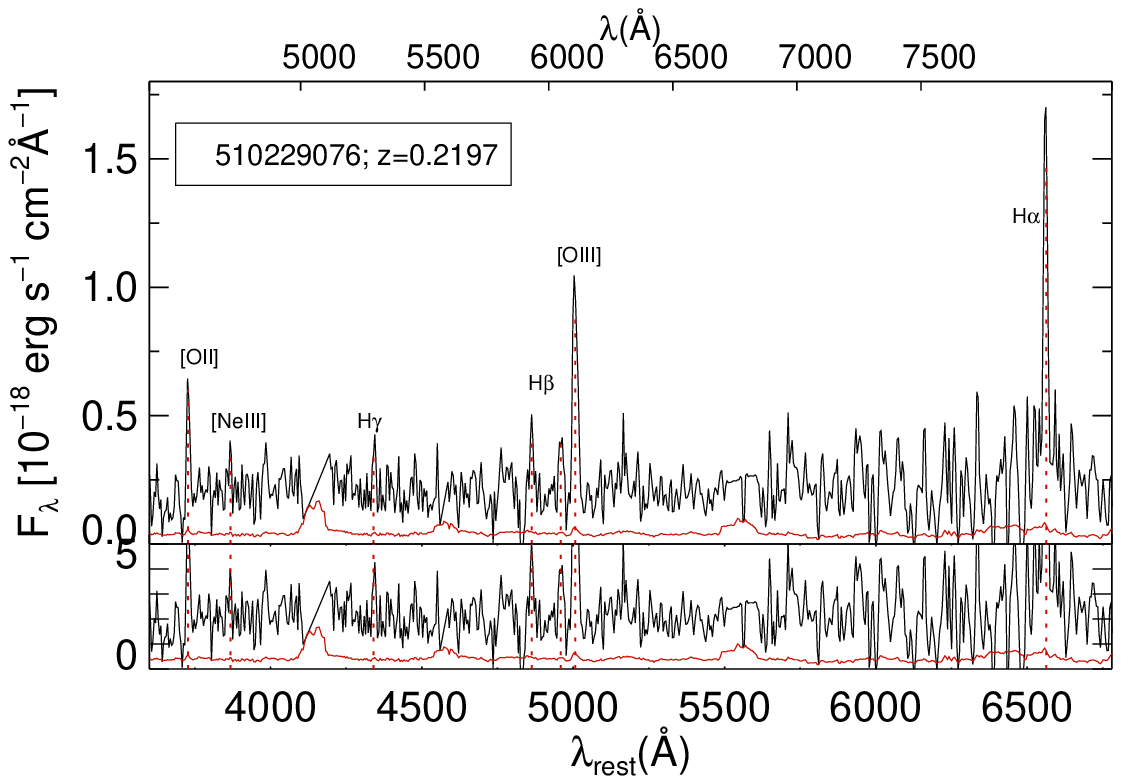} \\[0.5cm]
   \includegraphics[angle=0,width=8.5cm]{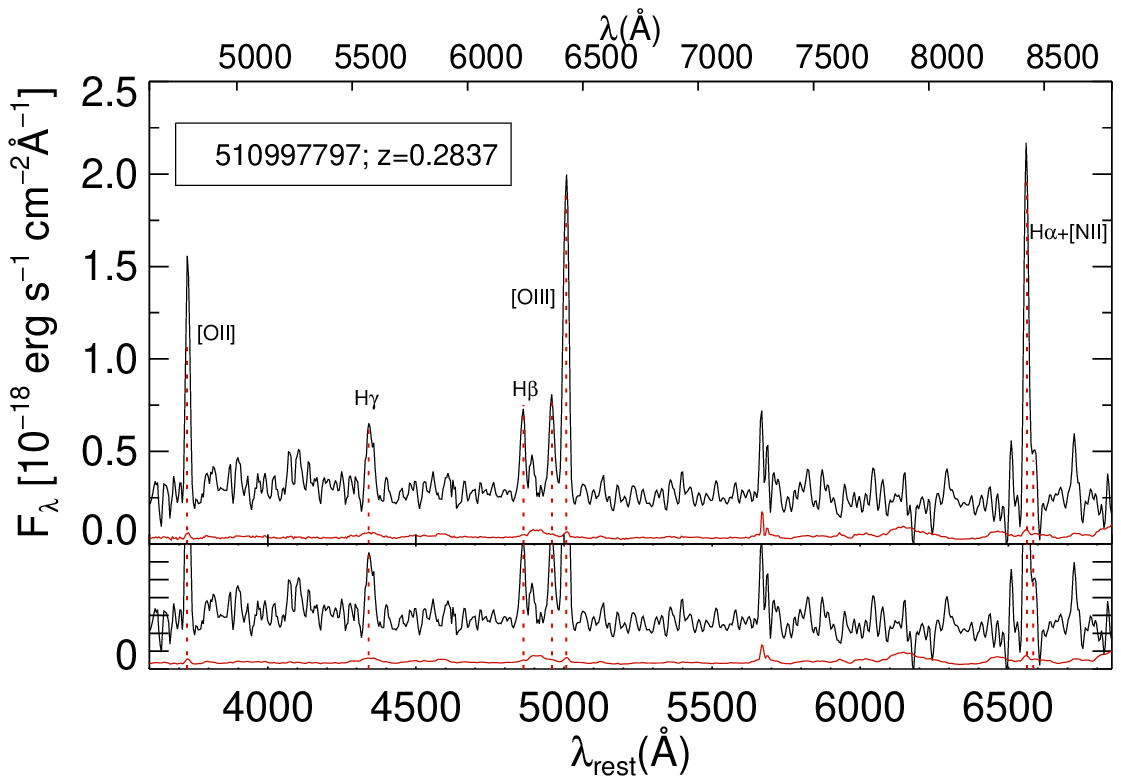} \hspace*{0.5cm}
   \includegraphics[angle=0,width=8.5cm]{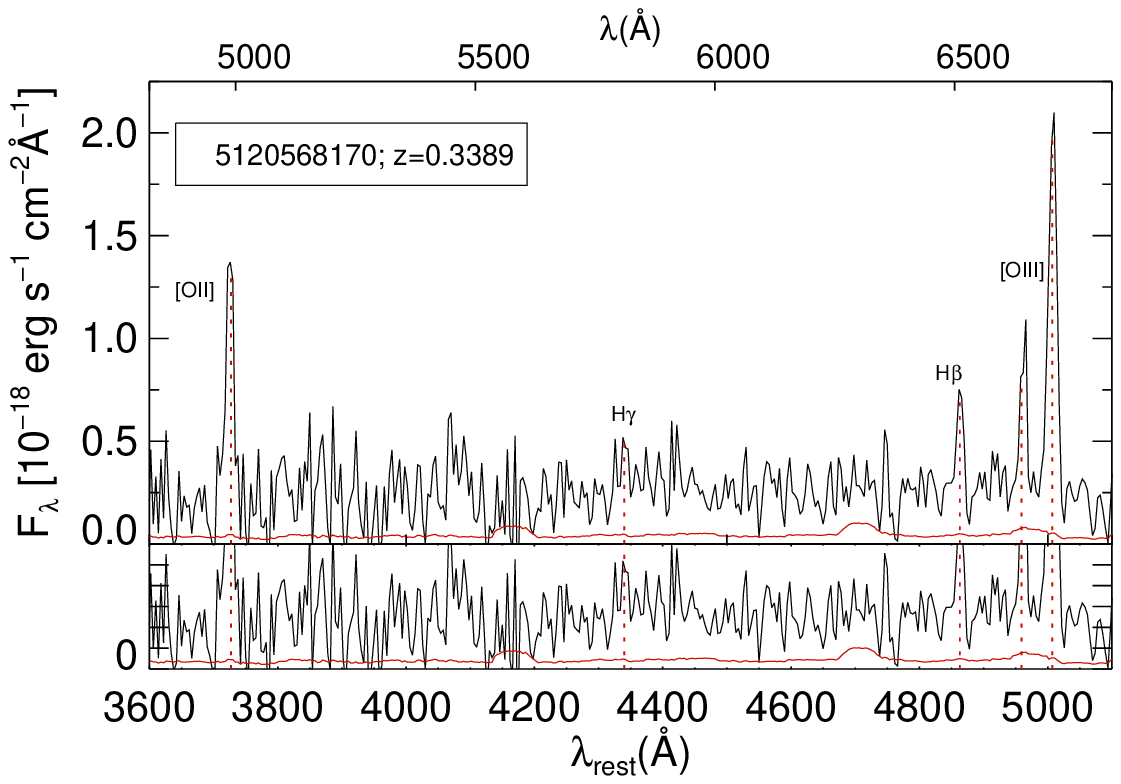} \\[0.5cm]
   \includegraphics[angle=0,width=8.5cm]{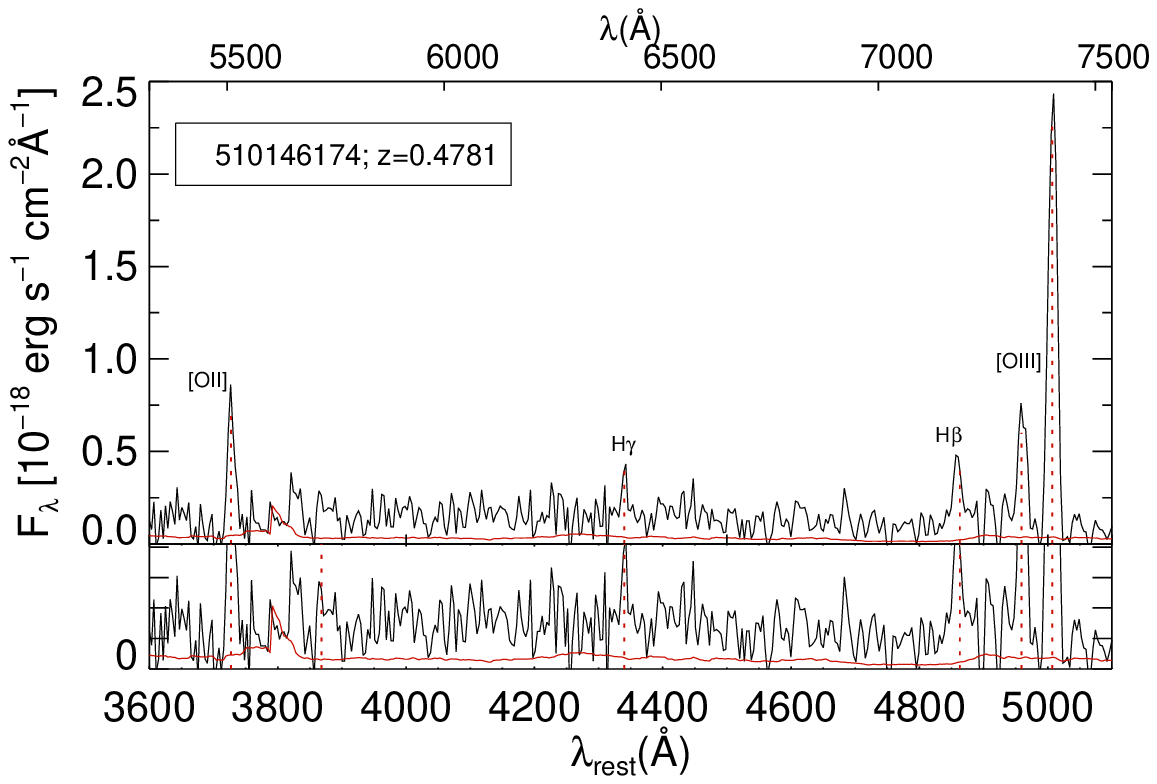} \hspace*{0.5cm}
   \includegraphics[angle=0,width=8.5cm]{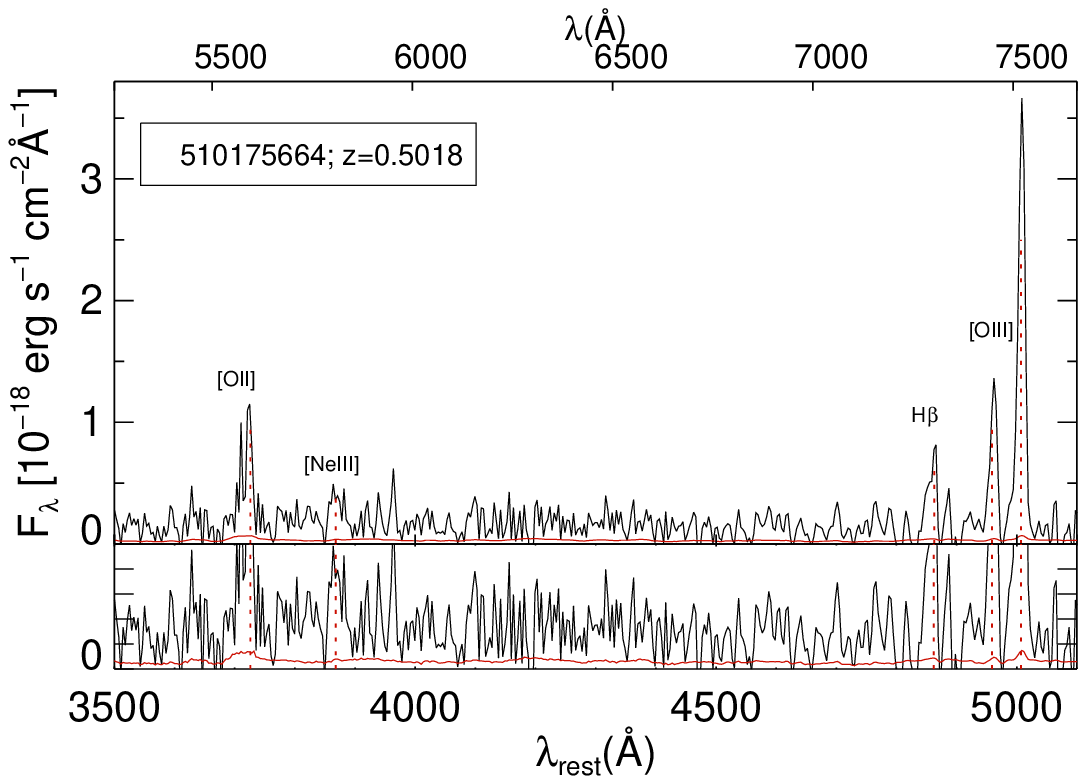} \\[0.5cm]
\caption[]{\label{atlas} VUDS spectra for the sample galaxies. Continued}
     \end{figure*}
 
    \begin{figure*}[t!] 
   \ContinuedFloat
   \includegraphics[angle=0,width=8.5cm]{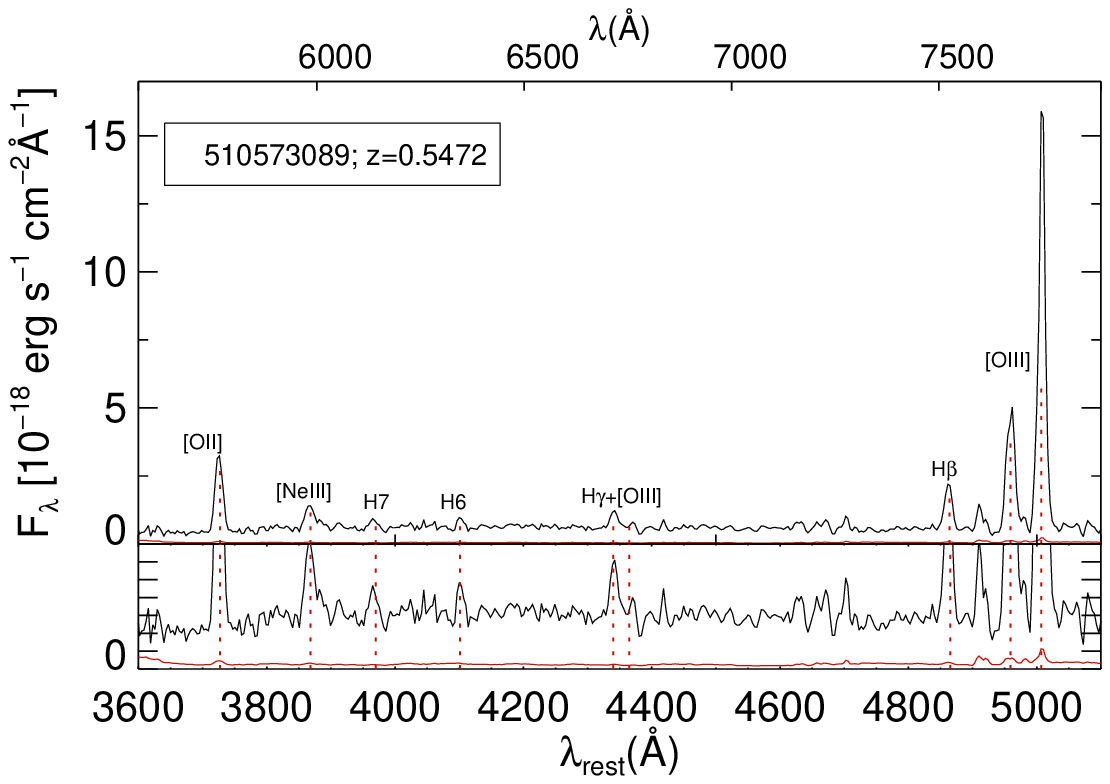} \hspace*{0.5cm}
   \includegraphics[angle=0,width=8.5cm]{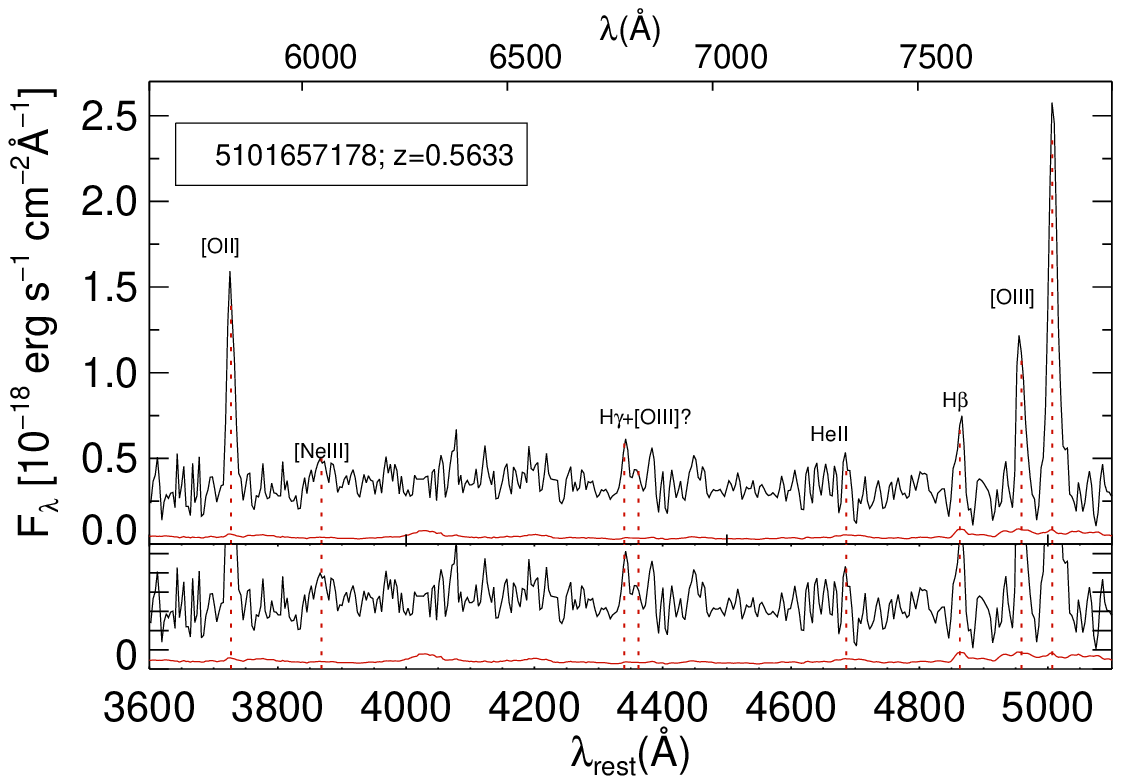} \\[0.5cm]
   \includegraphics[angle=0,width=8.5cm]{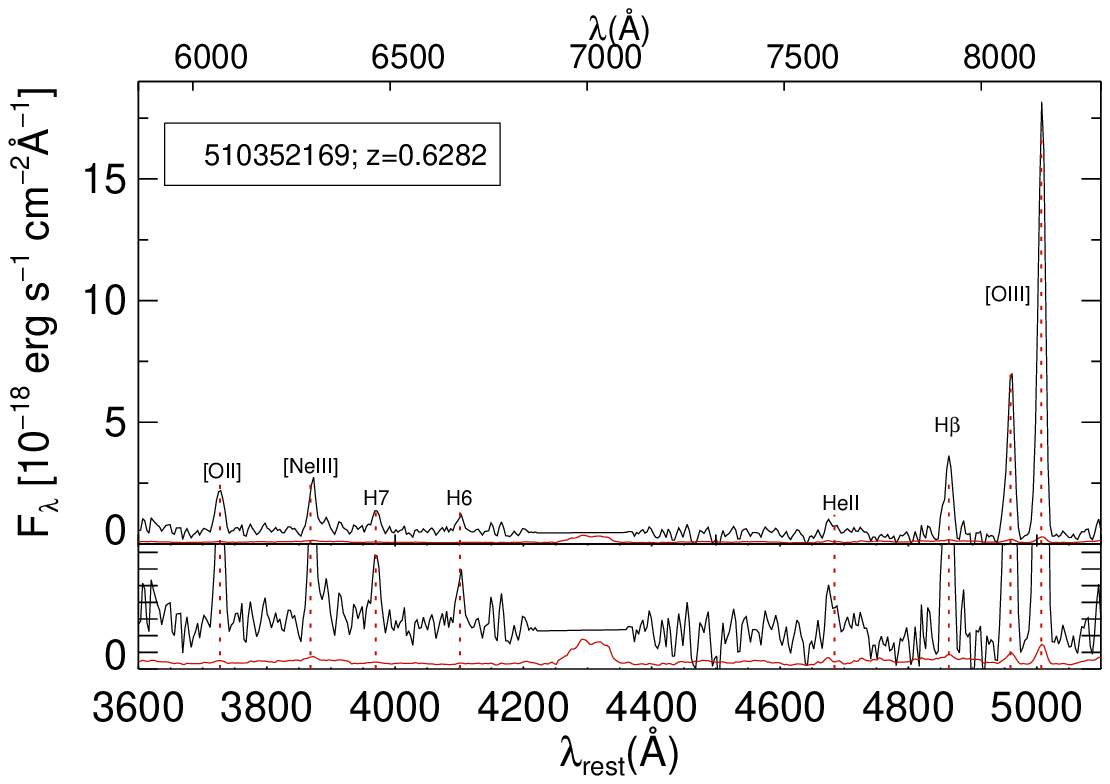} \hspace*{0.5cm}
   \includegraphics[angle=0,width=8.5cm]{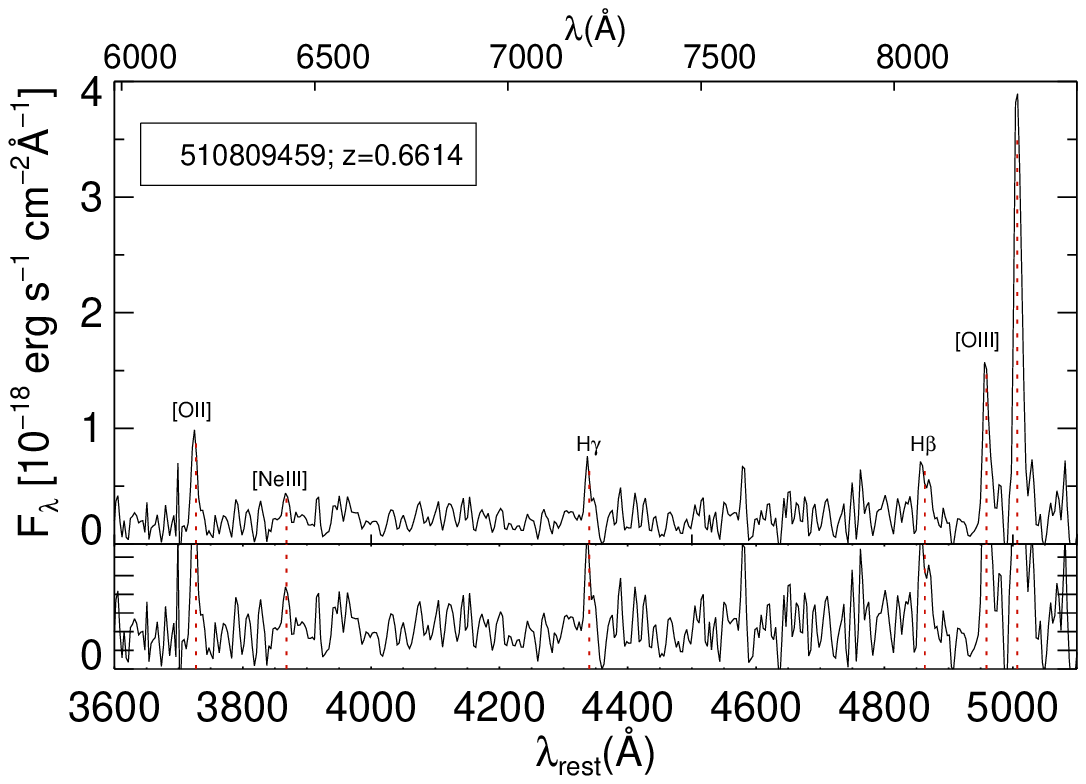} \\[0.5cm]
   \includegraphics[angle=0,width=8.5cm]{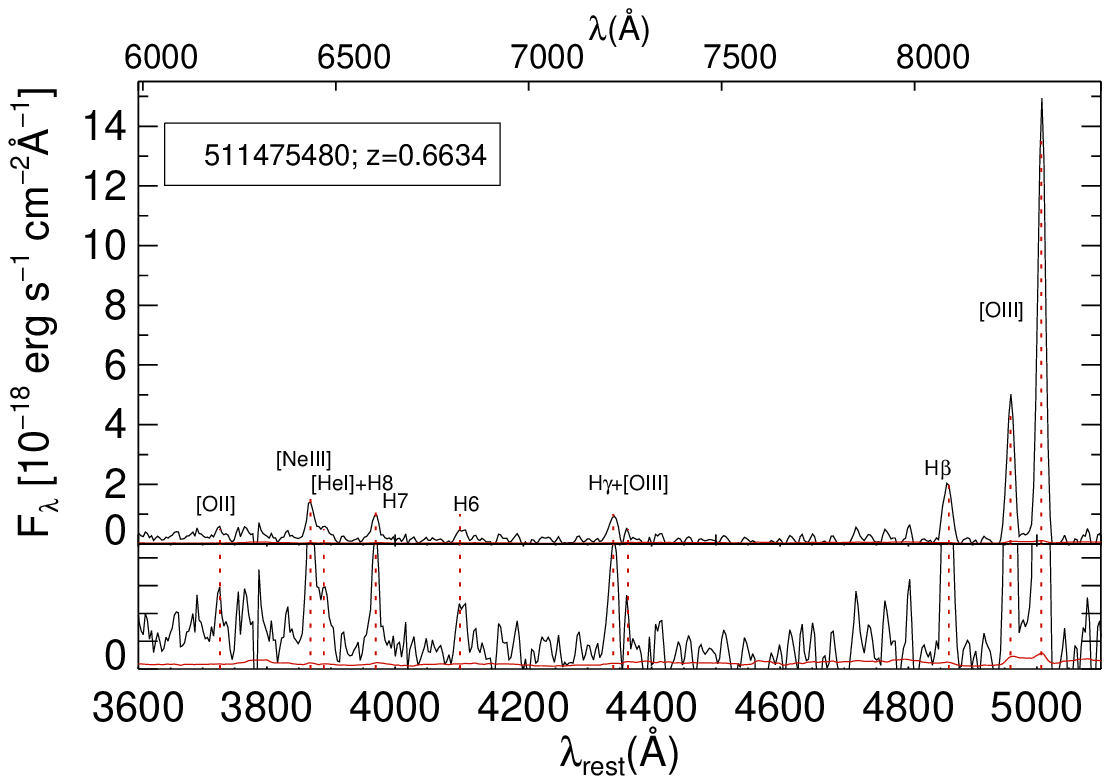} \hspace*{0.5cm}
   \includegraphics[angle=0,width=8.5cm]{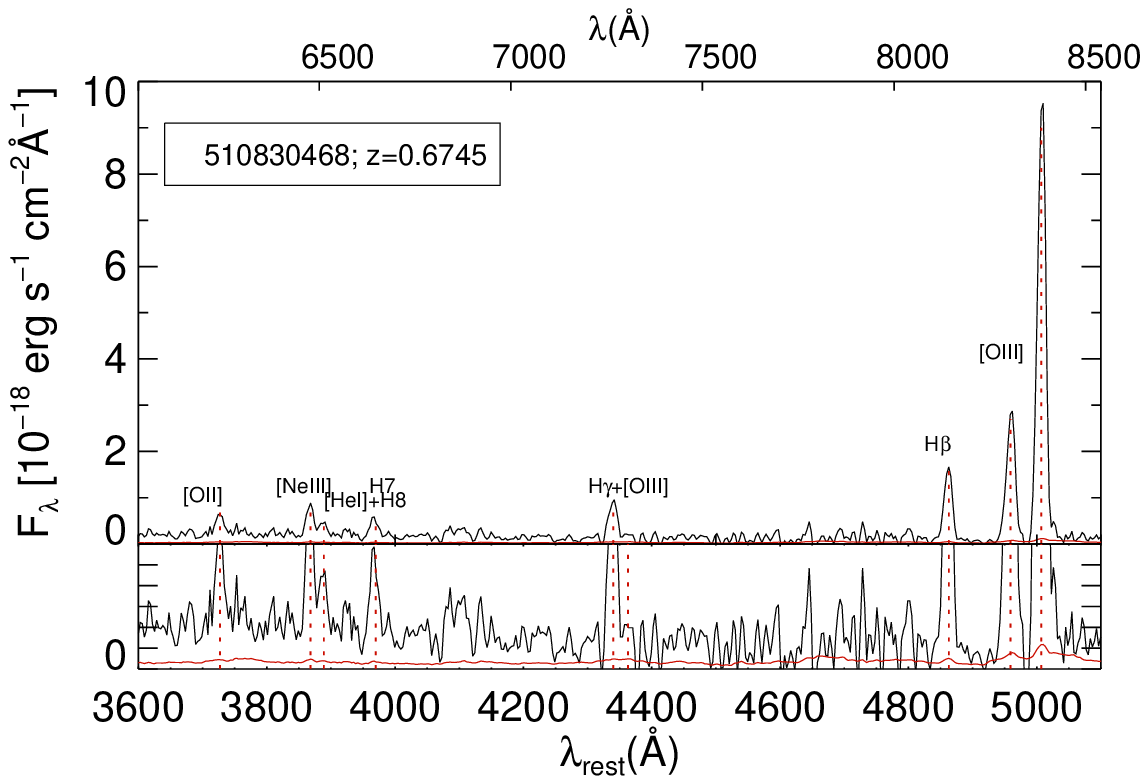} \\[0.5cm]
   \includegraphics[angle=0,width=8.3cm]{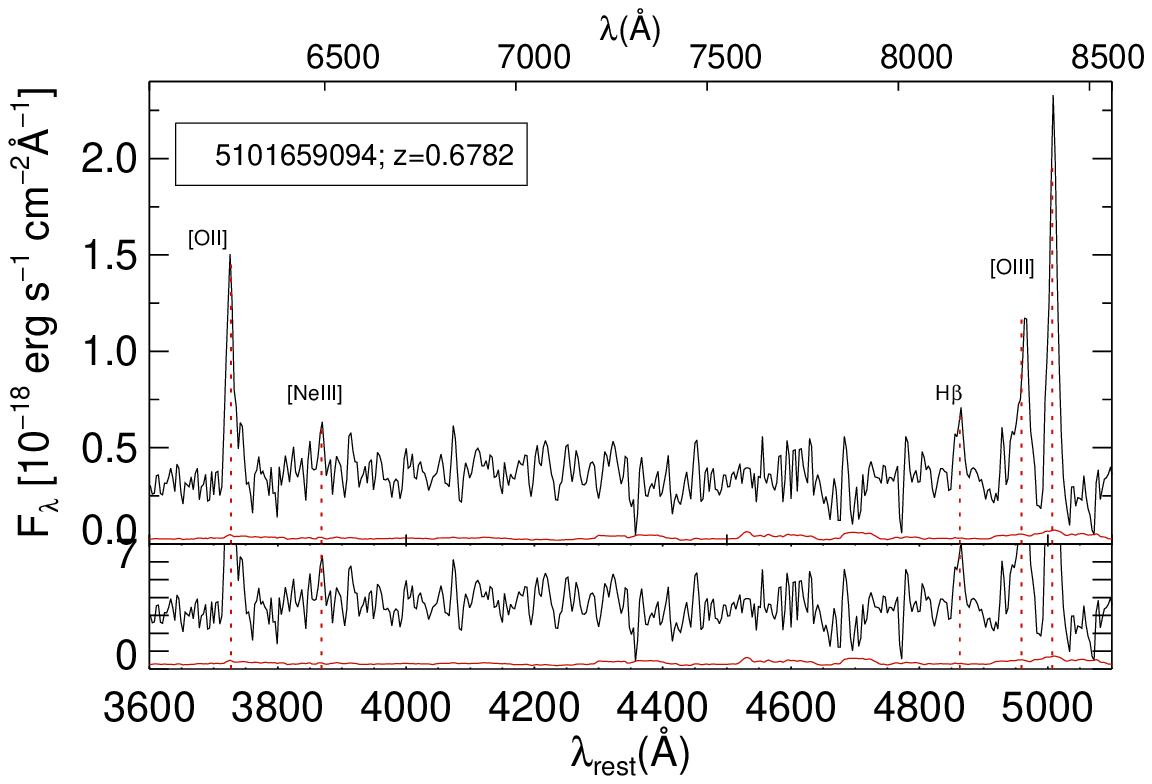} \hspace*{0.5cm}
 \caption[]{\label{atlas} VUDS spectra for the sample galaxies. Continued}
    \end{figure*}
 }
%-------------------------------------------------------------  
%\end{appendix}
%\clearpage
%\newpage
\clearpage

\end{document}